\documentclass{article}
\usepackage{epsf,epsfig,amssymb,float} 
\newcommand{\beq}{\begin{equation}}
\newcommand{\eeq}{\end{equation}}
\newcommand{\beqa}{\begin{eqnarray}}
\newcommand{\eeqa}{\end{eqnarray}}

\newcommand{\vs}{\vspace{-0.25cm}}

\begin{document}

\hfill {\tiny FZJ-IKP(Th)-2002-01}

\vspace{2cm}

\begin{center}

{{\Large\bf Chiral dynamics with strange quarks:\\[0.3em] 
            Mysteries and opportunities}}\footnote{Contribution to the Eta Physics Handbook,
    Workshop on Eta Physics, Uppsala 2001.}

\end{center}

\vspace{.3in}

\begin{center}
{\large Ulf-G. Mei{\ss}ner}\footnote{Email:~u.meissner@fz-juelich.de}

\bigskip

\bigskip

Forschungszentrum J\"ulich, Institut f{\"u}r Kernphysik (Theorie)\\
D-52425 J\"ulich, Germany

\end{center}

\vspace{0.5in}

\thispagestyle{empty}

\begin{abstract}
\noindent Recent developments and open issues in chiral dynamics with
strange quarks are reviewed. Topics include: Order parameters of chiral symmetry
breaking, the flavor dependence of these order parameters, speculations about
the phase structure of QCD with varying number of flavors, OZI violation and its
natural emergence in case of a small strange quark condensate, heavy kaon chiral
perturbation theory, an update on pion-pion, pion-kaon and pion-nucleon sigma terms,
the dynamics and nature of scalar mesons, the role of the scalar pion and
kaon form factors, attempts to describe some strange baryons employing
coupled channel dynamics tied to unitarity and chiral perturbation theory,
and the role of final state interactions in Goldstone boson pair production
in proton-proton collisions.
\end{abstract}

\vspace{5cm}

\centerline{PACS: 12.39.Fe, 14.40.Aq, 12.38.Aw, 11.30.Rd}

\vfill

\newpage

\section{Introduction}
\def\theequation{\arabic{section}.\arabic{equation}}
\setcounter{equation}{0}

The strange quark plays a special role in the QCD dynamics at the
confinement scale. In this essay, I will discuss some open questions
surrounding chiral dynamics with strange quarks, pertinent to the 
structure of the strong interaction vacuum as well as to the structure
of light mesons and baryons. Some of these questions are:
\begin{itemize}
\item[$\bullet$] Is the strange quark really light? It is well
  established that the up and the down quarks are light compared to any
  hadronic scale, in particular $m_u, m_d \ll \Lambda_{\rm QCD}$, with
  $\Lambda_{\rm QCD} \simeq 150\,$MeV.\footnote{If not specified
  differently, quark mass values refer to a scale of 1~GeV in the
  $\overline{\rm MS}$ scheme.} On the other hand, $m_s \sim  
  \Lambda_{\rm QCD}$ (the most recent non-lattice determination based
  on scalar sum rules gives $m_s (2\,{\rm GeV}) = 99 \pm 16\,$MeV
  \cite{JOP}. See that paper for many more references on
  determinations of $m_s$.) Therefore, one can even entertain the
  extreme possibility of considering the strange quark as heavy.
  Still, there are many successes of standard three flavor chiral
  perturbation theory based on the conventional scenario which treats the
  strange quark mass as a perturbation on equal footing with the light
  quark masses $m_u$ and $m_d$ (see e.g. the contributions of
  Ametller \cite{Aeta}, Bijnens \cite{Beta}, Gasser \cite{Geta} 
  and Holstein \cite{Heta} to this volume). 
\item[$\bullet$] Why is the OZI rule so badly violated in the scalar
   sector with vacuum quantum numbers? This rule is exact in the limit
   of a large number of colors $N_c \to \infty$ and works
   astonishingly well in most channels even for $N_c =3$, i.e. in 
   nature. However, in the $0^{++}$ channel, huge deviations from
   the OZI rule are observed.  One example is the reaction
   $J/\Psi \to \phi \pi\pi /\bar{K} K$, which is OZI suppressed to
   leading  order (see Fig.~\ref{fig:OZI}a) but has an additional
   doubly OZI suppressed contribution depicted in Fig.~\ref{fig:OZI}b.
   These processes have been measured at DM2 and Mark-III. The $\pi^+
   \pi^-$ event distribution shows a clear peak at the energy of
   980~MeV which is due to
   the $f_0$ scalar meson. This lets one anticipate that the dynamics
   of the low-lying scalar mesons and the mechanism of OZI violation
   are in some way related.
   \begin{figure}[H]
   \centerline{
   \epsfysize=1.7 in 
   \epsffile{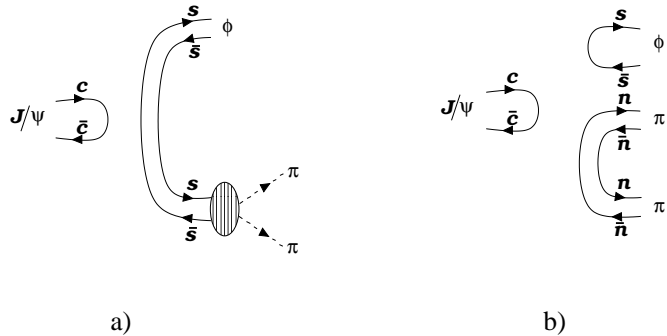}
   }
   \vskip .1cm 
   \centerline{\parbox{10cm}
   {\caption{Quark line diagrams for decay $J/\Psi$ into the $\phi$ and a meson
   pair ($\pi\pi$ or $K\bar{K}$). The quark flavors are explicitly given,
   $n$ refers to the light non--strange $u,d$ quarks. The hatched blob in a)
   depicts the final state interactions in the coupled $\pi\pi/K\bar{K}$ system.
   \label{fig:OZI}}}}
   \end{figure}
\item[$\bullet$] More generally, it is of interest to learn about the
  phase structure of SU($N_c$) gauge theory at large number of flavors
  $N_f\,$. In QCD, we know that asymptotic freedom is lost for $N_f
  \geq 17$ but from the study of the two-loop $\beta$ function  one
  expects that there is a conformal window around $N_f \simeq 6$, see
  \cite{BaZa} and for a recent update with many references, see
  \cite{GG}. This lets one contemplate the question whether there is already
   a rich phase structure even for the transition from $N_f =2$ to
  $N_f =3$~? Some lattice studies seem to indicate a strong flavor
  sensitivity when going from $N_f =2$ to $N_f =4$. In
  Fig.~\ref{fig:lat} I show some results obtained by the Columbia group
  using staggered fermions. These seem to indicate a suppression of
  the chiral condensate as the number of flavors is increased together
  with the degeneracy of meson and baryon states of opposite parity
  \cite{Col}. For similar lattice results, see \cite{aoki}.
   \begin{figure}[htb]
   \centerline{
   \epsfysize=4cm 
   \epsffile{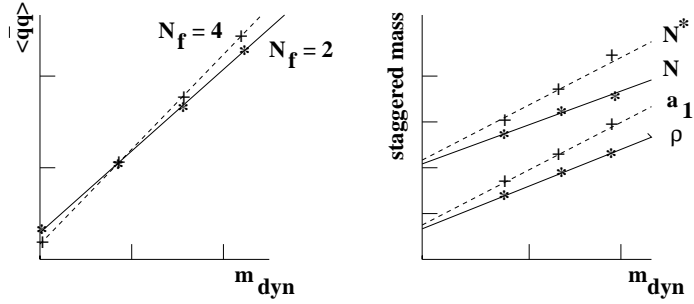}
   }
   \vskip .1cm 
   \centerline{\parbox{10cm}
   {\caption{Left panel: Suppression of the chiral condensate with
   increasing number of flavors. Right panel: Degeneracy of staggered
   hadron masses for $N_f =4$ which seems to indicate a restoration of
   chiral symmetry. From the QCDSP collaboration.
   \label{fig:lat}}}}
   \end{figure}   
\item[$\bullet$] The so-called sigma terms play a special role in the
  investigations of chiral dynamics chiefly because they are nothing
  but the expectation values of the quark mass term of QCD within
  hadron states, $H$, $\langle H| m_q \bar{q}q | H \rangle$, with $H$
  e.g.  pions, kaons or nucleons. Since no external scalar probes
   are available, the determination of these matrix elements proceeds
   by analyzing four--point functions, more precisely Goldstone boson--hadron 
   scattering amplitudes in the unphysical region, $\phi (q) + H(p)
  \to \phi (q') + H (p')$ (note that the hadron can also be a
  Goldstone boson). While the pion and kaon sigma terms behave as
  expected, there is still much debate about the pion-nucleon sigma
  term, especially concerning the possible admixture of $\bar{s} s$ pairs
  in the proton's wave function (which itself can, of course, not be
  measured). This might be taken as a hint that the transition from
  two to three flavors reveals some unexpected behavior. 
\item[$\bullet$] The nature of the low-lying scalar mesons, the
 so-called ``sigma'', the $f_0 (980)$, $a_0 (980)$ and so on, is still
 under debate -  are these quark-antiquark states, kaon-antikaon
 molecules or dynamically generated by the strong final state
 interactions in the coupled channel $\pi\pi / \bar{K}K$ system?
 Is there furthermore a mixing of some of these states with glueballs~?
 To my opinion, any model to describe these states must be able to
 describe the large body of data on decays (strong and
 electromagnetic) as well as scattering and production reactions.
 Also, since these particles have vacuum quantum numbers, they play a 
 special role as we already encountered in the brief discussion of OZI
 violation  and the phase structure of QCD for more than two flavors. 
\item[$\bullet$] In the baryon sector, there are also some ``strange''
 states with non-vanishing strangeness. More precisely, what is the
 nature  of some strange baryons like the $\Lambda (1405)$
 or the $S_{11} (1535)$, are these three quarks states or meson-baryon bound
 states~? The latter scenario was already contemplated many years ago 
 by Dalitz and Tuan \cite{Dal} and has been rejuvenated with the advent of
 coupled channel calculations using chiral Lagrangians to specify the
 driving interaction. The bound-state scenario is triggered by the fact that
 the $\Lambda (1405)$ sits just above/below the $\pi \Sigma /
 \bar{K}N$ threshold as shown in Fig.~\ref{fig:splane}.
   \begin{figure}[htb]
   \centerline{
   \epsfysize=3cm 
   \epsffile{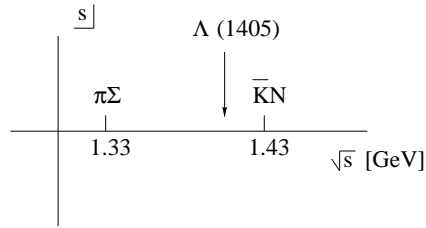}
   }
   \vskip .1cm 
   \centerline{\parbox{7cm}
   {\caption{Location of the $\Lambda (1405)$  between the $\pi
   \Sigma$ and the $\bar{K}N$ thresholds.
   \label{fig:splane}}}}
   \end{figure}
\end{itemize}

\noindent This essay is organized as follows. In
section~\ref{sec:chisym} I briefly discuss the chiral symmetry of QCD and
its breaking with particular emphasis on the pertinent order
parameters. Section~\ref{sec:exact} discusses QCD inequalities for 
these order parameters as a function of $N_f$, the number of light
quark flavors. Further facts and speculations about the flavor
dependence of the quark condensate are collected in
section~\ref{sec:cond}. The reordering of the chiral expansion in
light of a possible suppression of the three flavor quark condensate
is discussed in section~\ref{sec:reorder}. Another type of reordering
based on the assumption that one can treat kaons and etas has heavy
particles is presented in section~\ref{sec:heavyK} and some pertinent
results are discussed. Then, in section~\ref{sec:sigma} the status of 
sigma terms as extracted from pion-pion, pion-kaon and pion-nucleon
scattering is reviewed.  Some remarks on the structure of the light
scalar mesons and certain strange baryons are made in
sections~\ref{sec:scalar} and \ref{sec:baryon}, respectively. Finally,
I will make some comments about pseudoscalar (strange) meson production in
proton-proton collisions in section~\ref{sec:pp}. The essay concludes 
with a short summary and outlook in section~\ref{sec:summ}.

\section{Chiral symmetry of QCD - order parameters}
\def\theequation{\arabic{section}.\arabic{equation}}
\setcounter{equation}{0}
\label{sec:chisym}

The QCD Lagrangian has the form
\beq
{\cal L}_{\rm QCD} = -\frac{1}{4} G_{\mu\nu}^\alpha G^{\mu\nu , \alpha} - 
\frac{g^2}{32\pi^2} {\theta} \tilde{G}_{\mu\nu}^\alpha {G}^{\mu\nu , \alpha}
+ \sum_q \bar{q} (i D \!\!\!\!/ - m_q)q + \sum_Q \bar{Q} (i D\!\!\!\!/ - M_Q)Q 
+ {\cal L}_{\rm gf}~,
\eeq
with $G_{\mu\nu}^\alpha$ the conventional gluon field strength tensor, 
$\tilde{G}_{\mu\nu}^\alpha$ its dual, the color index $\alpha$ runs from 1 to 8,
 $g$ denotes the strong coupling constant,
$D\!\!\!\!/ = \gamma^\mu D_\mu =   \gamma^\mu (\partial_\mu -i g 
G_\mu^\alpha T^\alpha)$ the gauge-covariant derivative 
(with $T^\alpha = \lambda^\alpha/2$ the SU(3) generators),
$q$ collects the light quark fields
$u$, $d$, and $s$ whereas the heavy quarks $c$, $b$ and $t$ are collected
in $Q$. These fields are chosen such that the current quark mass matrix
is diagonal, its entries are the three real numbers $m_q$ and the three $M_Q$. 
The topological term involving the dual field strength tensor leads to
strong $P$ and $T$ violation. Experimental bounds like from the electric
dipole moment of the neutron, however, force the vacuum angle $\theta$
to be very tiny. Therefore, in what follows, we set the $\theta$ angle to zero. 
The term ${\cal L}_{\rm gf}$ collects gauge--fixing and ghost terms.
The heavy quark sector is governed by the so-called heavy quark symmetry,
which allows for many interesting predictions. The light quark sector, on the
other hand, is dominated by the so-called {\em chiral symmetry}, as will be
explained shortly. We will consider the heavy quarks as infinitely heavy, that
is they decouple from the light quark sector. Of course, there are interesting
processes which show an intricate interplay between heavy quark and chiral
symmetry, but we will not be concerned with these here. To a good first
approximation, one can set the light quark masses to zero, $m_q = 0$.
In that case the quark fields can be decomposed into right- and left-handed
parts, which do not interact. Stated differently, one can independently rotate
the right- and the left-handed light quark fields,
\beq
q_{R/L} \to V_{R/L}\, q_{R/L}~, \quad  V_{R/L} \in {\rm SU}(3)_{R/L}~.
\eeq
Obviously, a fermion mass term like $\bar{q}_L {\cal M} q_R + 
\bar{q}_R {\cal M}^\dagger q_L$ breaks this chiral symmetry, but it can
still be considered an approximate symmetry if the quark masses are in some
sense light. Chiral symmetry leads to eight conserved right- and eight
left-handed currents. Alternatively, one consider eight conserved vector
($V = L+R$) and eight conserved axial-vector ($A = L-R$) currents and the
corresponding conserved charges:
\beq
\begin{array}{llllll}
\partial^\mu V_\mu^a &=& i \bar{q} [ {\cal M}, \lambda^a ] q
& \rightarrow & Q_V^a = \int_V d^3 V_0^a 
& {\rm with} ~~\displaystyle{{dQ_V^a \over dt}} =0~,  \\ \nonumber \\
\partial^\mu A_\mu^a &=& i \bar{q} \{ {\cal M}, \lambda^a \}\gamma_5 q
& \rightarrow & Q_A^a = \int_V d^3 V_0^a
& {\rm with} ~~\displaystyle{{dQ_A^a \over dt}} =0~,  
\end{array}
\eeq
with $a=1,\ldots,8$ and ${\cal M} = {\rm diag}(m_u,m_d,m_s)$ the
diagonal current quark matrix. In three flavor {\em chiral limit},
that is $m_u =m_d = m_s = 0$, these 16 currents and charges are
exactly conserved. It is believed that QCD undergoes 
spontaneous chiral symmetry breaking and that the symmetry is realized
in the Nambu-Goldstone mode, i.e.
\beq
\lim_{V \to \infty} \langle 0 | [ {\cal O}, Q_V^a ] | 0 \rangle = 0 ~,
\quad {\rm but:} \quad 
\lim_{V \to \infty} \langle 0 | [ {\cal O}, Q_A^a ] | 0 \rangle \neq 0 ~, 
\eeq
where ${\cal O}$ is some order parameter (as discussed in more detail below)
and $| 0 \rangle$ denotes the highly complicated strong interaction vacuum.
It is one of the few exact theorems of QCD that in the absence of $\theta$-terms,
the vector symmetry cannot be broken (more generally this holds for any vector-like 
gauge theory under the stated assumptions)
\cite{VaWi}. There is also ample phenomenological evidence that the axial
symmetry is not realized in nature. For example, the hadron spectrum reveals
SU(3)$_V$ multiplets (the celebrated eightfold way), but no parity doublets,
as it would be the case for an additional realized axial symmetry, are observed.
The axial symmetry is hidden, its dynamics is transfered into the interactions
of the Goldstone bosons, which are a consequence of the symmetry violation. 
In massless QCD, there are eight of these modes and they should be massless.
Collectively, I denote these degrees of freedom as {\em  pions}. As a direct 
consequence of Goldstone's theorem, these pions can couple to the vacuum via the
axial current,
\beq
 \langle 0 | A_\mu^a (x) | \pi^b(p) \rangle = i \delta^{ab} F(3) p_\mu 
{\rm e}^{i p \cdot x}~,
\eeq
with $F(3)$ the pion decay constant in the three flavor chiral limit, i.e.
\beq
\lim_{m_u,m_d,m_s \to 0} F_\pi = F(3)~.
\eeq
In the real world, that is massive QCD, these Goldstone bosons can be 
identified with the three pions ($\pi^+, \pi^-, \pi^0$), the four kaons
($K^+, K^-, K^0, \bar{K}^0$) and the eta ($\eta$). They are the lightest
hadrons and their finite mass is related to the small current quark masses.
We will come back to this topic in section~\ref{sec:cond}. 

\begin{figure}[htb]
\centerline{
\epsfysize=5.0cm
\epsffile{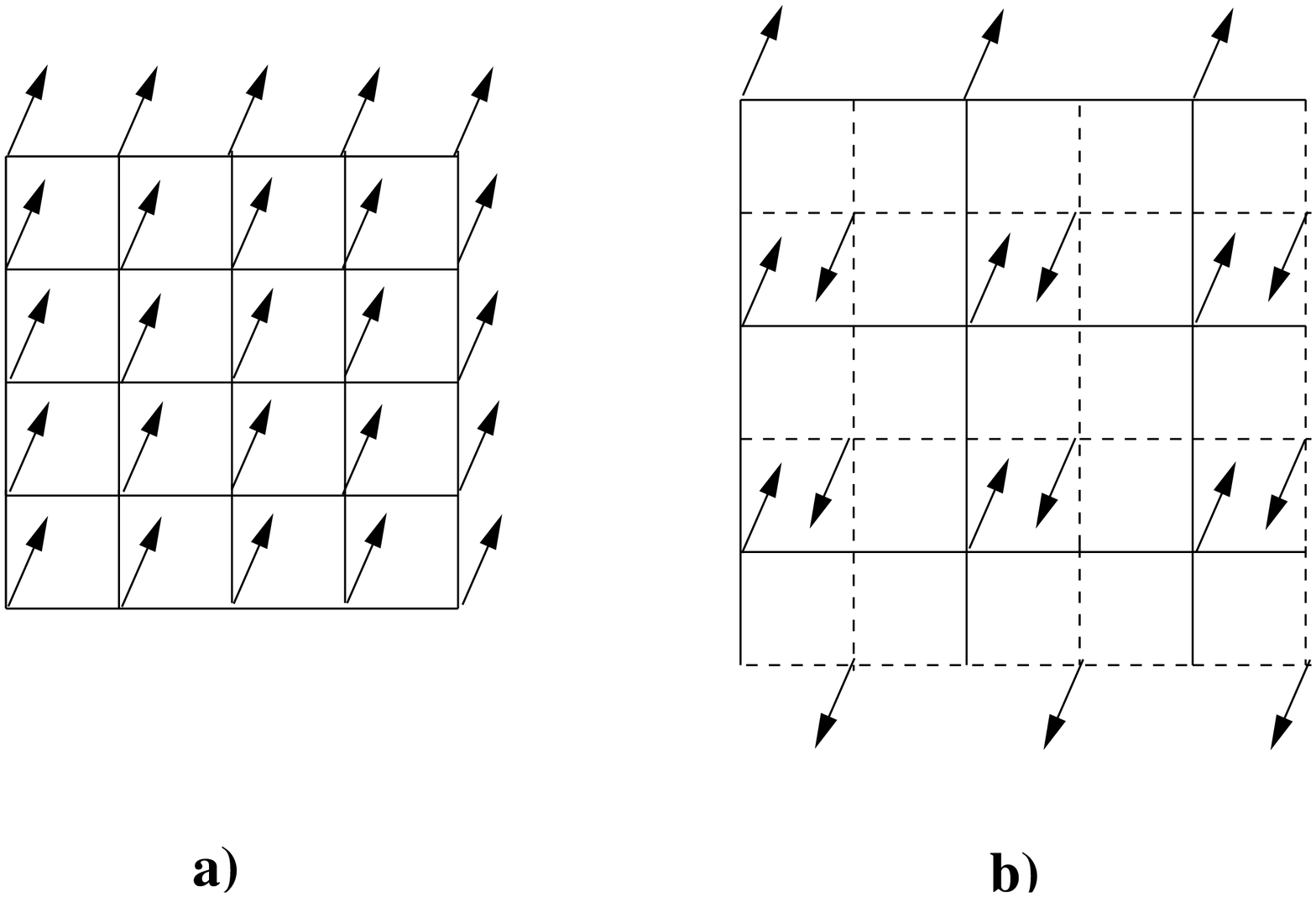}
}
\centerline{\parbox{10cm}
{\caption{a) Ferromagnetic order. The spins on all lattice sites point
          in the same direction. This defines the magnetization.
          b) Anti--ferromagnetic order. There are
          two sublattices as shown by the solid and dashed lines. The spins
          on the sites of these sublattices point in opposite directions,
          such that the total magnetization vanishes.
         \label{fig:order}}}}        
\end{figure}

\medskip\noindent
Let me return to the notion of an order parameter. A textbook example is a
magnetic system of spins located on some lattice sites interacting with 
an energy $\sim \vec{S}_i \cdot \vec{S}_j$, where $i$ denotes some site
and $j$ belongs to its nearest neighbors. Obviously, the magnet Hamiltonian
is invariant under spatial rotations, i.e under O(3). However, below the
Curie temperature this symmetry is spontaneously broken,
O(3)~$\to$~O(2), with
all spins pointing in one direction as shown in panel a) of Fig.~\ref{fig:order}. 
\noindent
This defines the magnetization $\vec{M}$, and one has $\langle 0| \vec{M} |0 \rangle
\neq 0$, where $|0 \rangle$ denotes the ground state. The remaining O(2) symmetry
group is given by the rotations around the direction of the magnetization.
More generally, one defines an order parameter ${\cal O}$ by the relation:
\beq\label{order}
\langle 0| {\cal O} |0 \rangle \neq 0 \quad \rightarrow \quad
{\rm spontaneous}~~{\rm symmetry}~~{\rm breaking}~.
\eeq
For the ferromagnet, the magnetization is the simplest order parameter, in fact, there
is an infinite amount of such order parameters. For the ferromagnetic case,
any positive power of $\vec{M}$ qualifies as such.
The inverse of Eq.~(\ref{order}) is, however, not true. This can
be  easily exemplified using an anti-ferromagnetic system, as shown
in the panel b) of Fig.~\ref{fig:order}. Here, the total magnetization
is zero, because there are two sublattices, on which the spins point in
opposite direction, as shown in the figure by the dashed and the solid
lines. The pertinent order parameter in this case is $ \vec{M}_{\rm solid}-
\vec{M}_{\rm dashed}$, where the magnetization is defined on the corresponding
sublattices. More complicated order parameters, which account for this underlying
structure, can of course be considered. From this simple example one concludes
that a detailed study of the order parameters in a given system 
can reveal the nature of the mechanism of the spontaneous symmetry breaking.

\medskip\noindent
Now the question arises what are the order parameters of the spontaneous
chiral symmetry breaking\ ? Consider first the current-current correlator
between vector and axial currents,
\beqa
\Pi^{ab}_{\mu\nu} (q) &=& i \int d^4x \, {\rm e}^{i q x} \, 
\langle 0| T\{ V_\mu^a (x)  V_\nu^b (0) -  A_\mu^a (x)  A_\nu^b (0) \} |0\rangle
\nonumber \\
 &=& 4i \int d^4x \, {\rm e}^{i q x} \, \langle 0| T \{R_\mu^a (x)  L_\nu^b (0)\}
 |0\rangle~.
\eeqa
In the three flavor chiral limit, it can be written in terms of meson
and continuum contributions, i.e. for $m_q \to 0$ we have
\beq
\Pi^{ab}_{\mu\nu} (q) = (q_\mu q_\nu -g_{\mu\nu} q^2)  
\, \left[ -\sum_P {F^2(3) \over q^2}
- \sum_V {F_V^2 \over M_V^2 -q^2} + \sum_A {F_A^2 \over M_A^2 -q^2}
 + \ldots \right] ~, 
\eeq 
where the first sum runs over the pseudoscalar Goldstone bosons,
the second over vector and the third over axial-vector mesons.
It follows immediately that
\beq
\Pi^{ab}_{\mu\nu} (0) = -\frac{1}{4} g_{\mu\nu} \delta^{ab} F^2(3)~.
\eeq
We have  thus identified an order parameter of spontaneous chiral 
symmetry breaking (S$\chi$SB), namely the pion decay constant in the
chiral limit. Its non-vanishing is a sufficient and necessary condition
for S$\chi$SB,
\beq
\Pi^{ab}_{\mu\nu} (0) \neq 0 \leftrightarrow F(3) \neq 0 
 \leftrightarrow {\rm  S}\chi{\rm SB}~.
\eeq
Naturally, there are many other possible order parameters, often
considered is the light quark condensate,
\beq\label{qbarq}
\langle 0| \bar{q} q |0\rangle = \langle 0| \bar{u} u |0\rangle^{(3)} 
= \langle 0| \bar{d} d |0\rangle^{(3)} = \langle 0| \bar{s} s |0\rangle^{(3)}
\equiv -\Sigma(3)~,
\eeq 
because the scalar-isoscalar operator $\bar{q}q$ mixes right- and left-handed
quark fields. As will be discussed in the following sections, the quark condensate
plays a different role than the pion decay constant. Other possible color-neutral
order parameters of higher dimension are e.g. the mixed quark--gluon condensate 
$\langle 0| \bar{q}^i \sigma^{\mu\nu} G_{\mu\nu}^\alpha T^\alpha_{ij} q_j 
|0\rangle^{(3)}$ or certain four-quark condensates  $\langle 0| (\bar{q} \Gamma_1 q)
 (\bar{q} \Gamma_2 q) |0\rangle $ with $\Gamma_i$ some Dirac operator. This concludes
our general discussion of order parameters in QCD. It goes without saying that
the spontaneously and explicitely broken chiral symmetry can be systematically 
analyzed in terms of an effective field theory - chiral perturbation theory
(CHPT) (or some variant thereof) \cite{We79,GL1,GL2}.

\section{QCD inequalities: Flavor dependence of chiral order parameters}
\def\theequation{\arabic{section}.\arabic{equation}}
\setcounter{equation}{0}
\label{sec:exact}

We now consider some exact results for the flavor dependence of order
parameters of the S$\chi$SB in QCD. For this, we consider QCD on a torus,
or in an Euclidean ($t \to -ix^0$) box of size $L\times L \times L \times L$, 
understanding of course that we have to take the infinite volume limit $L \to \infty$ 
at its appropriate place.  Quarks and gluon fields are then subject certain
boundary conditions, which are anti-periodic and periodic, in order.
I will not go through the derivation of these results, but
rather refer to the work of Banks and Casher \cite{BaCa}, Vafa and Witten
\cite{VaWi}, Leutwyler and Smilga \cite{LeSm}, Stern \cite{Jan} and
Descotes, Girlanda and Stern \cite{DGS} (and others). The interested reader
is referred to these articles for the details.

\medskip\noindent
To be specific, consider the Dirac operator of QCD in a gluonic background,
\beq
{\cal D} [G] = \gamma_\mu ( \partial_\mu + i G_\mu ) = {\cal D}^\dagger [G]~,
\eeq
where the important hermiticity property is rooted in the Euclidean metric
we are using. The eigenvalue equation for the Euclidean Dirac operator reads
\beq
{\cal D} \!\!\!\!/ \, \psi_n = \lambda_n \, \psi_n~.
\eeq
From this, three important observations can be made. First, the spectrum
of ${\cal D}[G]$ is symmetric under $\psi_n \to \gamma_5 \psi_n$, that is
we can classify the  eigenvalues $\lambda_n \geq 0$ by a strictly positive,
increasing number (since $\lambda_{-n} = -\lambda_n$). In addition, 
for gauge fields with non-vanishing winding number $\nu$, the spectrum 
contains $|\nu |$ topological zero modes. Second,
one can derive an uniform upper bound for the small eigenvalues \cite{VaWi},
\beq\label{bound}
| \lambda_n [G] | < C \, {n^{1/d} \over L} \equiv \omega_n~,
\eeq
where $d$ is the dimensionality of space-time and the constant $C$ does
not depend on the gauge field configuration, the integer $n$ and the 
volume $V=L^d$ but only on the shape of the space-time manifold (at
fixed volume).  Third, only the small eigenvalues are relevant to generate S$\chi$SB
in the infinite volume limit (see below).  Consider now some operator for varying number
of flavors, more specifically for the transition $N_f \to N_f + 1$. 
The flavor dependence of such an operator enters through the quark propagator,
\beq 
S(x,y|G) = \sum_n {\psi_n(x) \psi_n^\dagger (y) \over {\cal M} + i \lambda_n}~,
\eeq
and through the fermion determinant. More precisely, consider first the quark 
condensate for $N_f$ flavors with a uniform light mass $m$ in a gluonic background,
\beqa\label{BC1}
\Sigma (N_f) =  \lim_{L \to \infty} {1 \over L^4} \left\langle \int
  d^4x \, {\rm Tr} \, S(x,x|G)\right\rangle_{[G]}^{N_f} 
&=& \lim_{L \to \infty} {1 \over L^4} \left\langle \sum_n
{\psi_n(x) \psi_n^\dagger(x) \over m + i \lambda_n }\right\rangle_{[G]}^{N_f}
\nonumber \\
&=&   \lim_{L \to \infty} {1 \over L^4} \left\langle \sum_n {m \over
m^2 + \lambda_n^2} \right\rangle_{[G]}^{N_f} \nonumber\\
&=& \lim_{m \to 0} \int_{-\infty}^{+\infty} d \epsilon {m \over m^2 + \epsilon^2}
\,  \lim_{L \to \infty}  {1 \over L^4} 
 \Biggl\langle \underbrace{\sum_n \delta (\epsilon -
     \lambda_n)}_{\equiv \rho (\epsilon, L )} \Biggr\rangle_{[G]}^{N_f}~,
\eeqa
where the symbol $\langle~\rangle_{[G]}^{N_f}$ represents the
normalized average over gauge field configurations weighted by the
fermion determinant,
\beq\label{av}
\left\langle O \right\rangle_{[G]}^{N_f} = Z^{-1} \int d\mu [G] \, O\,
\Delta^{N_f} (m|G) \, \prod_{j>N_f} \Delta (m_j|G) \, \exp(-S[G])~.
\eeq
Here, $S[G]$ stands for the Yang-Mills action and $Z$ ia a
normalization constant. This functional
integral may also be viewed as an average over all Dirac spectra.
Note that Eq.~(\ref{BC1}) is nothing but the Banks-Casher relation \cite{BaCa}
\beq\label{BC}
\Sigma (N_f) = \pi \, \lim_{\epsilon \to 0} \lim_{L \to \infty} 
\rho (\epsilon, L)~,
\eeq
which expresses the quark condensate in terms of the zero modes
of the QCD Dirac operator. From this formula one concludes that
an accumulation of small eigenvalues $|\lambda_n| \sim 1/L^4$ is necessary
so that $\Sigma (N_f) \neq 0$. A similar relation can be obtained
for the pion decay constant \cite{Jan}
\beq
F^2 (N_f) =  \lim_{L \to \infty} {1 \over L^4} \left\langle \sum_{k,n}
{m \over m^2 + \lambda_k^2}{m \over m^2 + \lambda_n^2} J_{kn}
 \right\rangle_{[G]}^{N_f}~,
\eeq
in terms of the ``quark mobility''
\beq
J_{kn} = \frac{1}{4} \sum_\mu \left| \int d^4x\, \psi^\dagger_n (x) \gamma_\mu
\psi_n (x) \right|^2~.
\eeq
This can be brought in a form similar to the Banks-Casher relation, Eq.~(\ref{BC}),
\beq
F^2 (N_f) = \pi^2 \, \lim_{\epsilon \to 0} \lim_{L \to \infty} 
L^4 \, J(\epsilon, L) \, \rho^2 (\epsilon, L)~.
\eeq
Clearly, an accumulation of  small eigenvalues $|\lambda_n| \sim 1/L^2$ is 
necessary so that $F^2 \neq 0$. We note the different IR sensitivity as
compared to the quark condensate. Now we return to Eq.~(\ref{av}). The
existence of the bound Eq.~(\ref{bound}) allows to split the fermion
determinant into IR and UV parts \cite{DET} by choosing a cut-off
$\Lambda$ and an integer $K$ such that $\omega_K = \Lambda$. The
single flavor fermion determinant takes the form
\beq
\Delta (m|G) = m^{|\nu|} \, \Delta_{\rm IR} (m|G)\, \Delta_{\rm UV}
(m|G)~,
\eeq
where $\Delta_{\rm IR}$ involves the first $K$ non-zero eigenvalues
and it  is bounded by one,
\beq
 \Delta_{\rm IR} (m|G) = \prod_{k=1}^K {m^2 + \lambda_k^2 [G] \over m^2
   + \omega_k^2} < 1~.
\eeq
Consequently, since the order parameters $F^2$ and $\Sigma$ are
dominated by the IR end of the Dirac spectrum, the flavor dependence
resides mostly in the determinant $\Delta_{\rm IR}$ and one therefore
expects a paramagnetic effect,
\beqa
\Sigma (N_f+1) &<& \Sigma (N_f) ~~\sim 1/L^4~, \\
F^2 (N_f+1) &<& F^2  (N_f) ~\sim 1/L^2~,
\eeqa
indicating a suppression of the chiral order parameters with
increasing number of flavors.
We stress again that the condensate is most IR sensitive. These results are
exact, the question is now how strong this flavor dependence is or how
this flavor dependence can be tested or extracted from some observables.
This will be the topic of the next section.

\section{Flavor dependence of the quark condensate: Facts and\\ speculations}
\def\theequation{\arabic{section}.\arabic{equation}}
\setcounter{equation}{0}
\label{sec:cond}

We are now in the position to analyze the flavor dependence of the quark
condensate. First, consider the {\em standard scenario}, in which
$\Sigma(3) \lesssim \Sigma(2)$. With the definition of the quark
condensate given in Eq.~(\ref{qbarq}) the quark mass expansion of the
Goldstone bosons takes the form (neglecting intra-multiplet mass splittings
and $\pi^0-\eta$ mixing)
\beqa\label{mass}
F_\pi^2 M_\pi^2 &=& 2 \hat{m} \, \Sigma(3) + {\cal O}(m_q^2)~, \nonumber\\
F_K^2 M_K^2 &=& (\hat{m} + m_s) \, \Sigma(3) + {\cal O}(m_q^2)~, \nonumber\\
F_\eta^2 M_\eta^2 &=& \frac{2}{3}(\hat{m}+2m_s) \, \Sigma(3) + {\cal O}(m_q^2)~,
\eeqa
with $\hat{m} = (m_u+m_d)/2$ the average (two-flavor) light quark mass.
Note that the differences between $F_\pi$, $F_K$ and $F_\eta$ are formally
of higher order in the chiral expansion. However, for later comparison
we chose to work with this particular form of writing the Goldstone
boson masses in terms of the quark masses.
In the standard scenario, the terms quadratic in the quark masses are
small, as has been recently confirmed for  the {\em two flavor} case
from the analysis of the BNL E865 $K_{e4}$ data \cite{E865}. If these terms
are also small in the three flavor case, the so-called Gell-Mann--Oakes--Renner
ratio $X(3)$ stays close to one,
\beq 
X(3) \equiv { 2 \hat{m} \Sigma (3) \over F_\pi^2 M_\pi^2} \sim 1~.
\eeq
If the terms quadratic in the quark masses are exactly zero, the
celebrated Gell-Mann-Okubo relation results:
\beq\label{GMO}
M_\eta^2 = \frac{4}{3}M_K^2 - \frac{1}{3}M_\pi^2~,
\eeq
which is fulfilled in nature within a few percent, $(547.3\,{\rm
  MeV})^2 \simeq (569.5\,{\rm MeV})^2$.
This is one of the strong arguments in favor of the standard scenario,
since corrections to the   Gell-Mann-Okubo relation are small. Note that
to leading order in the quark mass expansion, QCD is characterized by
two parameters (apart from the quark masses), namely $F(3)$ (of dimension
energy) and $\Sigma(3)$ (of dimension energy cubed).
In the standard case, the corresponding scales are very different,
\beq
\Sigma^{1/3} (3) \sim 1.5~{\rm GeV} \gg F(3) \sim 0.1~{\rm GeV}~.
\eeq
For the application of this scheme to processes involving pions, kaons and,
especially, etas, I refer to the contributions by  Ametller, Bijnens, Gasser 
and Holstein to this volume. Some related aspects concerning pion-kaon
scattering will be discussed in section~\ref{sec:heavyK}.

\begin{figure}[htb]
\centerline{
\epsfysize=5.0cm
\epsffile{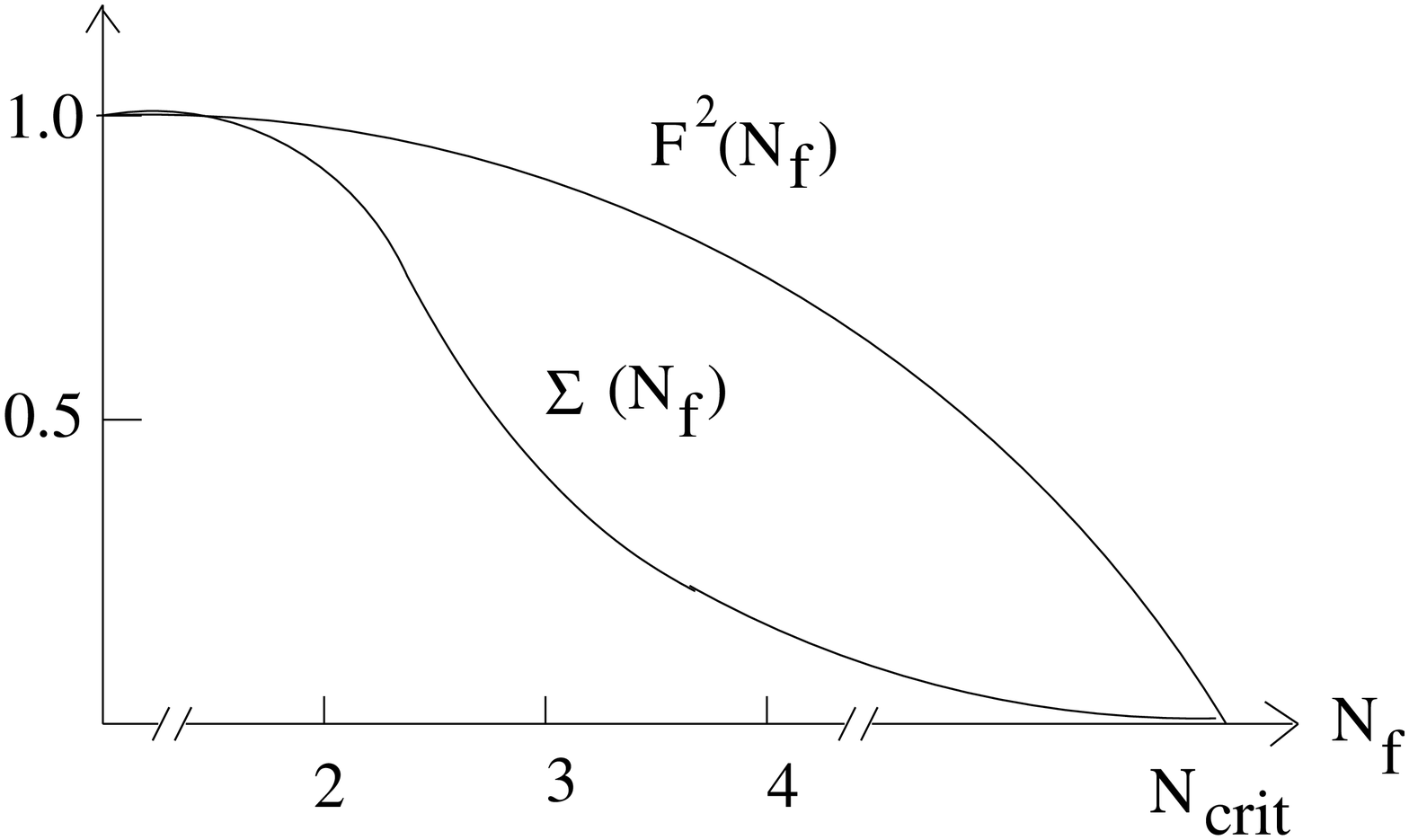}
}
\centerline{\parbox{10cm}
{\caption{Possible non--standard phase structure of QCD for increasing number 
      of flavors $N_f$. Shown are two (arbitrarily normalized) order parameters.
      While the pion decay constant $F^2$ varies only mildly, the quark condensate
      $\Sigma$ drops sharply when $N_f$ goes from two to three.
      \label{fig:phase}}}}        
\end{figure}

\begin{figure}[ht]
\parbox{.49\textwidth}{\epsfig{file= 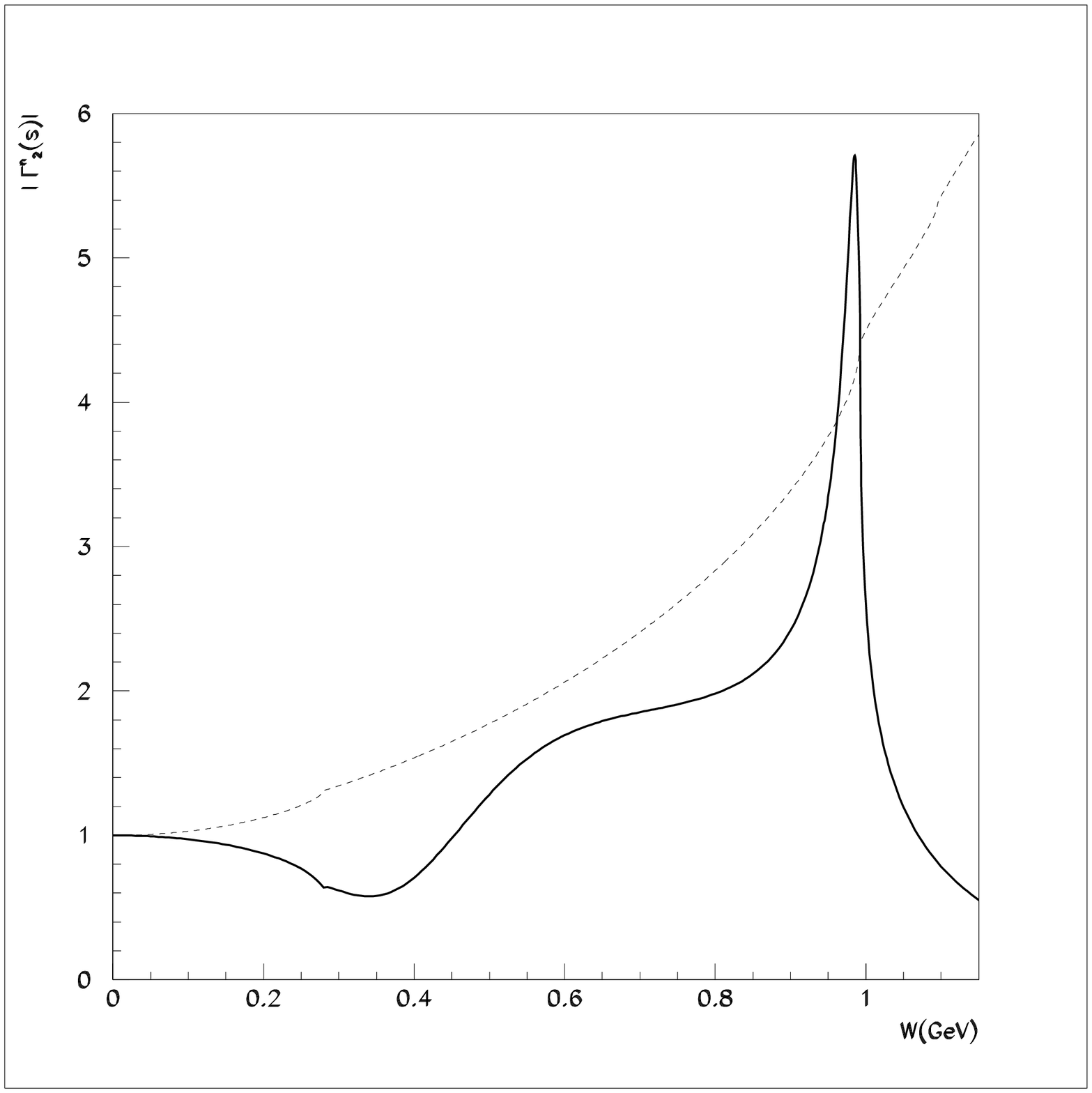,width=.47\textwidth,silent=,clip=}}
\hfill
\parbox{.49\textwidth}{\epsfig{file= 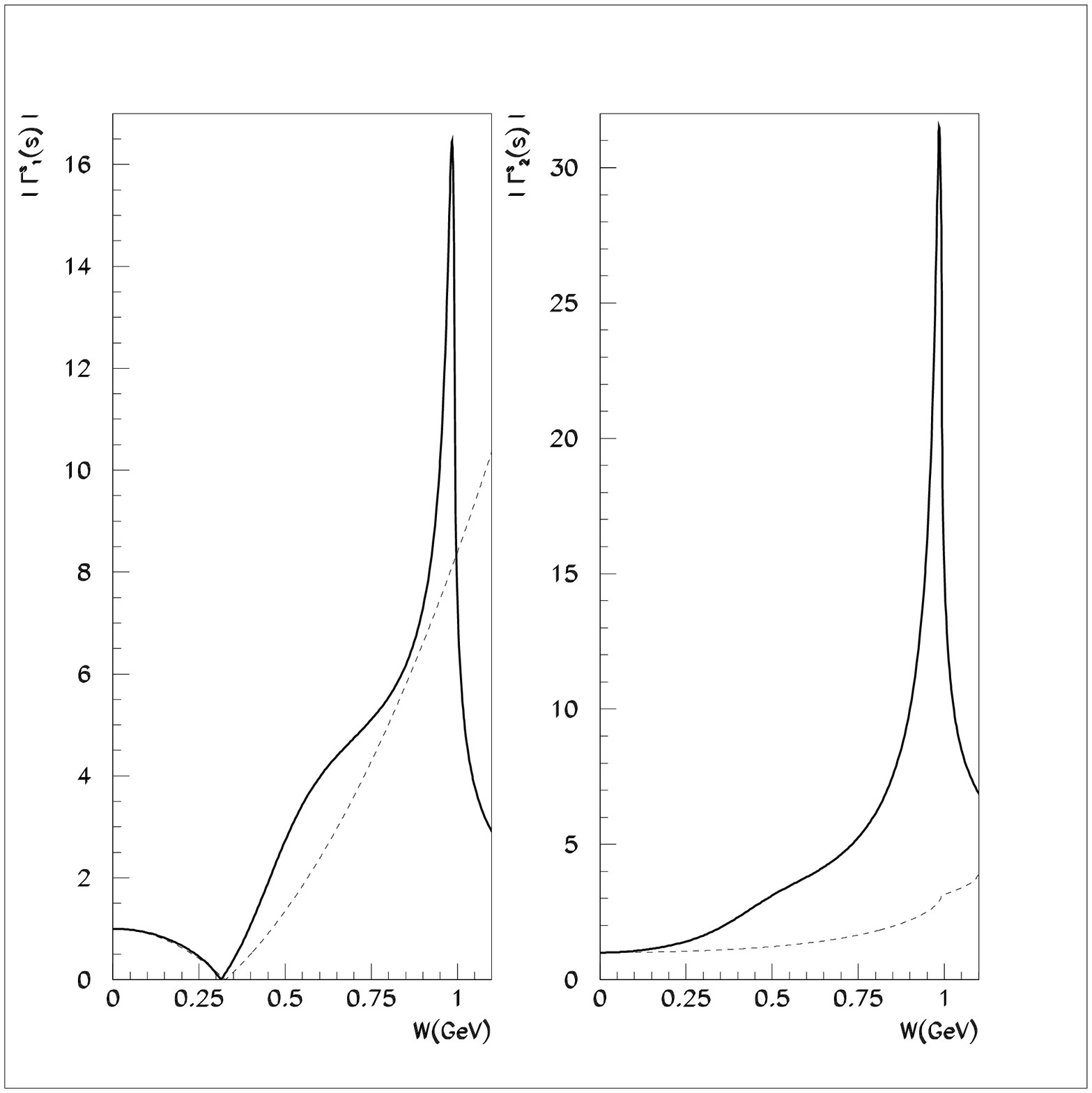,width=.47\textwidth,silent=,clip=}}
\caption{\label{fig:sff}
           Scalar form factors as a  
           function of the energy $W = \sqrt{s}$ in GeV. 
           Left panel: Modulus of the non-strange scalar kaon form 
           factor, $|\Gamma_K^n (s)|$. Right panel: Modulus of the strange 
           scalar form factor of the pion, $|\Gamma_\pi^s (s)|$, and
           the kaon,  
           $|\Gamma_K^s (s)|$, in order. Solid lines: Chiral unitary
           approach of \protect \cite{MeOl}, dotted lines: CHPT to 
           one-loop accuracy. The peak at $W \simeq 1\ $GeV is due
           to the $f_0 (980)$.
}
\end{figure}

\begin{figure}[H]
\centerline{
\epsfysize=6.5cm
\epsffile{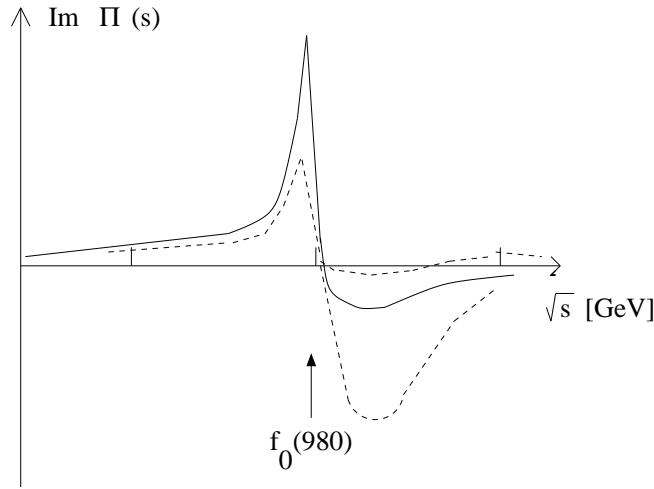}
}
\vspace{0.3cm}

\centerline{\parbox{10cm}
{\caption{Generic form of the spectral function $\sigma (s) = {\rm
      Im}~\Pi (s)$ as given by the solid line. The peak due to the 
      $f_0 (980)$ is clearly visible. The dashed lines indicate the
      uncertainty due to the various input $\pi \pi/ \bar{K}K$
      T-matrices.
      \label{fig:spec}}}}        
\end{figure}

\medskip\noindent
Let us now contemplate the possibility that $\Sigma (3) \ll \Sigma (2)\ $.
In an extreme scenario, this would mean that chiral symmetry is broken signaled
by $F(3) \neq 0$ but the condensate is very small, $\Sigma (3) \simeq 0$. 
In that case, we would have to consider a very different phase structure
of the QCD vacuum as it is the case of the standard scenario, where one
expands around a flavor symmetric phase, see Fig.~\ref{fig:phase}.
\noindent
To be able to test or analyze
such a scenario, we have to be more specific. Thus, consider the a theory
with $N_f$ massless quarks ($m_u = m_d = \ldots =  m = 0$) and one 
(strange quark) with a fixed mass $m_s$. The variation of the condensate
with the ``heavy'' mass can  easily be evaluated:
\beqa\label{Pis}
{\partial \over \partial m_s} \Sigma (N_f) &=& \lim_{m \to 0}
\int dx \, \underbrace{
\langle \bar u u(x) \bar s s(0) \rangle_c
}_{\rm OZI \!\!\!\!/}
\equiv \Pi_z (m_s) \nonumber \\
&=&   \lim_{L \to \infty} {1 \over L^4} \left\langle \left( \sum_n {m \over
m^2 + \lambda_n^2} \right) \, \left( \sum_k {m_s \over
m_s^2 + \lambda_n^2} \right) \right\rangle_{[G]~c}^{N_f}
\eeqa
where $c$ refers to the connected piece. The operator on top of the brace
has vacuum quantum numbers and clearly is related to OZI breaking. In the
limit of the vanishing light quark masses, we can integrate Eq.~(\ref{Pis})
and deduce the sum rule
\beq
\Sigma (N_f) = \Sigma (N_f+1) + \int_0^{m_s} d \mu \, \Pi_z (\mu) =
\Sigma (N_f+1) + m_s Z_s (m_s) + {\cal O}(m_s^2 \log m_s^2)~,
\eeq
where $Z_s$ is an OZI violating coupling constant. Note also that
$\Pi_z$ is an order parameter of S$\chi$SB. From this, one deduces
that in the standard scenario the difference $\Sigma (N_f) - \Sigma (N_f+1)$
is small for two reasons: First, $m_s \ll  \Sigma (N_f+1)$ and, second,
in the limit of a large number of colors, $N_c \to \infty$, 
$\langle \bar u u(x) \bar s s(0) \rangle_c$ vanishes chiefly because in that
limit the OZI rule is exact \cite{Witten}. Stated differently, in that case
quantum fluctuations are very much suppressed and the flavor dependence of
the quark condensate is very weak. However, if $N_f +1$ happens to be close
to a critical value $N_f^{\rm crit}$ (like e.g. for the conformal window),
then one would expect large fluctuations in the density of states and 
consequently the OZI suppressed operators (or low-energy constants 
in the language of chiral perturbation theory) would be enhanced.
This is the scenario which has been advocated by the Orsay group as 
a possibility over the last few years, see e.g. \cite{DGS}. 
The phase structure of the two most important order parameters, namely the
pion decay constant and the quark condensate, is the one shown 
in Fig.~\ref{fig:phase}. The question arises
whether there is any evidence for such a scenario? Indeed, there is some,
as first shown by Moussallam \cite{BM1,BM2}. I will briefly review his
arguments here and refer to \cite{descT,DeSt} for a refined discussion.
It was pointed out in \cite{BM1} that $\Pi_z$ satisfies a superconvergent
sum rule,
\beq\label{SR}
\Pi_z (m_s) = {1 \over \pi} \int_0^\infty {dt \over t} \sigma (t)~.
\eeq
Furthermore, the spectral function $\sigma (t)$ collects OZI-violating
contributions,
\beq
\sigma (p^2) = \frac{1}{2} \sum_n (2\pi)^4 \delta^{(4)} (p-P_n)
\langle 0 | \bar u u | n\rangle \langle n | \bar s s |0\rangle~.
\eeq
Here, the sum runs over isoscalar two-particle states, $| n\rangle = 
|\pi \pi \rangle, |K \bar K \rangle, \ldots \ $ (neglecting four-particle
intermediate states, which is supported by phenomenology) and the
matrix elements $\langle 0 | \bar u u | n\rangle = \langle 0 | \bar d d 
| n\rangle$ and  $\langle 0 | \bar s s | n\rangle$ are the so-called 
non-strange/strange scalar form factors of the pions and the kaons. 
More precisely:
\beq\label{sff} 
\begin{array}{ll}
\langle 0| \bar{n}n |\pi\pi\rangle = \sqrt{2} B_0 \Gamma_\pi^n (s)~, 
& \langle 0| \bar{n}n |\bar{K}K\rangle = \sqrt{2} B_0 \Gamma_K^n (s)~, \\ 
\langle 0| \bar{s}s |\pi\pi\rangle = \sqrt{2} B_0 \Gamma_\pi^s (s)~, 
& \langle 0| \bar{s}s |\bar{K}K\rangle = \sqrt{2} B_0 \Gamma_K^s (s)~, 
\end{array}
\eeq 
with $\bar{n}n = (\bar{u}u + \bar{d}d )/\sqrt{2}$ and $B_0$ characterizes the
strength of the quark-antiquark condensation in the nonperturbative vacuum, 
$B_0 = |\langle 0 |\bar{q}q|0\rangle|/F_\pi^2$. The superscript $n/s$ refers
to the non-strange/strange scalar-isoscalar operator. 
These form factors can be obtained from 
the coupled channel $\pi \pi \to \pi\pi / \bar{K}K$ T-matrix using CHPT constraints.
In  Fig.~\ref{fig:sff} I show some of these form factors taken from \cite{MeOl}.
%
\noindent
The strong peak due to the $f_0 (980)$ scalar mesons (which in the approach
used in \cite{MeOl} is dynamically generated by meson-meson final state interactions)
is also dominating the spectral function $\sigma (t)$. The generic form of this
spectral function is shown in Fig.~\ref{fig:spec}. In \cite{BM1} three
different T-matrices were used as input \cite{OOP,AMP,KLM}, 
they differ in the height due
to the $f_0$ meson and in particular for energies above 1~GeV,
fortunately such energies are suppressed due to the weight factor in
the sum rule Eq.~(\ref{SR}).
 In fact, for extracting $\Pi_z$, one decomposes the energy
integration into three regions. Region I extends up to $\sqrt{s}
\simeq 1.2\,$GeV and is dominated by the $f_0 (980)$ peak. Region II
must be mostly negative and can either be obtained from the explicit
T-matrices (as shown in Fig.~\ref{fig:spec}) or by another sum rule
\cite{Desc}. Finally, for energies above $1.6 \ldots 1.8\,$GeV, one
can use the OPE and duality since for large Euclidean momenta, $\Pi (p^2)
\to ((\alpha_s/\pi)^2 \times {\rm coefficient}/(-p^2))$. Putting all
this together, one arrives at $2 \lesssim 16\pi^2 \Pi_z \lesssim 6$
which translates into
\beq\label{Sig32}
\Sigma(3) = \Sigma (2) \, \left[ 1 - 0.54 \pm 0.27 \right]~,
\eeq
where the central value indicates a large suppression of the three
flavor condensate but the uncertainties are large enough to give
marginal consistency with the standard scenario. This result is stable 
against higher order corrections (for the scalar form factors) if one
assumes that the chiral series converges \cite{BM2}. Furthermore, one
can also analyze this in terms of the low-energy constants $L_i$ and
the quark mass ratio $m_s / \hat{m}$, as detailed in \cite{DeSt}\cite{Desc}.
Clearly, more work is needed to solidify these results and to reduce
the theoretical uncertainties. However, the central value of
Eq.~(\ref{Sig32}) is consistent with the assumption of a suppressed
three flavor condensate. Taken that for granted, I will discuss in the 
next section how the chiral expansion would have to be reordered.
Before doing that, it is important to stress that the possible
suppression of the three-flavor condensate, which is largely due to the
$f_0 (980)$ peak in the sum rule, leads also to a natural mechanism to 
explain the OZI violation in the sector with vacuum quantum numbers due
to the large light quark quantum fluctuations. It can also be tested
using lattice simulations for the small eigenvalues of the QCD Dirac
operator employing the extended Leutwyler-Smilga sum rules \cite{LeSm}
derived in \cite{DeSt2}. This is certainly an attractive
feature of this scheme, which needs to be explored in more detail. 
The $m_s$ dependence of the condensate has also been studied in \cite{ABT}.

\section{Reordering the chiral expansion I: Small condensate case}
\def\theequation{\arabic{section}.\arabic{equation}}
\setcounter{equation}{0}
\label{sec:reorder}

Let us now consider the case  $\Sigma (3) \ll \Sigma(2)$. Then, the
quark mass expansion of the Goldstone bosons is much more complicated than
it is in the standard case, Eq.~(\ref{mass}), because the leading term would
be suppressed and higher order terms in the quark masses could be equally
large. This scenario is, however, still predictive if one {\it assumes}
that the terms of third order in the quark mass only lead to small
corrections - as it would be expected if the couplings accompanying the
quadratic and cubic quark mass terms are all of natural size. Under this
assumption, the quark mass expansion of the Goldstone boson masses
takes the form\footnote{Note that this is more general than the so-called
generalized chiral perturbation theory in which the quark masses are 
counted as linear small parameters.} \cite{Desc}
\beqa\label{masss}
F_\pi^2 M_\pi^2 &=& 2 \hat{m} \Sigma(3) + 2\hat{m}(m_s+2\hat{m}) Z^s
+ 4\hat{m}^2 A + 4\hat{m}^2B_0^2 L + F_\pi^2 \delta_\pi~, \nonumber\\
F_K^2 M_K^2 &=& (\hat{m} + m_s) \Sigma(3) +
(m_s+\hat{m})(m_s+2\hat{m}) Z^s + (m_s+\hat{m})^2 A +
\hat{m}(m_s+\hat{m}) B_0^2 L + F_K^2 \delta_K~, \nonumber \\
F_\eta^2 M_\eta^2 &=& \frac{2}{3}(\hat{m}+2m_s) \Sigma(3) + 
\frac{2}{3}(\hat{m}+2m_s) (m_s +2\hat{m}) Z^s + \frac{4}{3}
(2m_s^2 + \hat{m}^2) A \nonumber \\
&& \qquad + \frac{8}{3} (m_s-\hat{m})^2 Z^p 
 +\frac{1}{3} B_0^2L + F^2_\eta \delta_\eta~,
\eeqa
where we define $B_0 = \Sigma (3) / F^2(3)$ and $L$ contains the
chiral logarithms,
\beq
L = {1\over 32\pi^2} \left[ 3 \log {M_K^2 \over M_\pi^2}  
+ \log {M_\eta^2 \over M_K^2} \right] = 25.3 \cdot 10^{-3}~,
\eeq 
which are, of course, dominated by the pion cloud.
The $\delta_P (P=\pi, K, \eta)$ are the assumed small remainders. 
The low-energy constants (LECs) $A$, $Z^s$ and $Z^p$ are related to the LECs 
$L_6$, $L_7$ and $L_8$ which appear at next-to-leading order in the 
standard scenario,  we have
\beqa
A &=& 16B_0^2 \left[L_8(\mu) - {1\over 512\pi^2} \left( \log
\frac{M_K^2}{\mu^2} + \frac{2}{3} \log\frac{M_\eta^2}{\mu^2} \right)\right]~,
\nonumber \\
Z^s &=& 32B_0^2 \left[L_6(\mu) - {1\over 512\pi^2} \left( \log
\frac{M_K^2}{\mu^2} + \frac{2}{9} \log\frac{M_\eta^2}{\mu^2}
\right)\right]~, \quad Z^p = 16B_0^2 \,L_7~.
\eeqa
If the $\delta_P$ are indeed small, the equations~(\ref{mass}) can be
analyzed in terms of the quark mass ratio $r = m_s / \hat{m}$ and the
LECs $L_i$. Before discussing these issues, it is important to reconsider
the Gell-Mann-Okubo relation in this context. In the way the meson decay
constants are included, it takes the somewhat unfamiliar form
\beq
3 F_\eta^2 M_\eta^2 + F^2_\pi M_\pi^2 = 4F_K^2 M_K^2~,
\eeq
which, of course, agrees with Eq.~(\ref{GMO}) since the differences between
the decay constants are ${\cal O}(m_q)$. For this equation to hold, one
must require
\beq
A + 2Z^p \simeq 0~,
\eeq
which, in terms of the standard LECs, amounts to a strong correlation
between $L_7$ and  $L_8$. Such a correlation remains to be explained, the
Gell-Mann-Okubo relation does not naturally arise here. Given now that
the remainders in Eq.~(\ref{masss}) are small, that is $\delta_P \ll
M_P^2$, one can derive a nonperturbative relation between the
three-flavor Gell-Mann-Oakes-Renner ratio $X(3)$, the quark mass
ratio $r = m_s/\hat{m}$ and the OZI-violating coupling $L_6$ (for a
given $F_0$, the pion decay constant in the chiral limit), see
\cite{DeSt}. Quantum fluctuations of the condensate, which scale as 
${\cal O}(1/N_c)$ are suppressed if $L_6$ lies in a narrow band around
$L_6 (M_\rho) \simeq -0.26 \cdot 10^{-3}$ (for $F_0 = 85\,$MeV), 
which is close to the value deduced in standard chiral
perturbation theory in connection with the OZI rule \cite{daphne}. However, the
dependence of $X(3)$ on $L_6$ is very strong, one finds e.g. $X(3)
\simeq 0.6$ for $L_6 (M_\rho) \simeq +0.4 \cdot 10^{-3}$ for $r \simeq
25$ and $F_0 = 85\,$MeV. Note, however, that the presence of three
massless flavors (in the sea) seems to be crucial for this effect -
vacuum fluctuations do not suppress the two-flavor condensate $X(2)$.
$X(2)$ remains close to one as long as $r>20$. In addition, using also
the quark mass expansions of $F_\pi$, $F_K$ and $F_\eta$, one can systematically
analyze the LECs $L_{4,5,7,8}$ as functions of $r$, $F_0$ and $L_6$.
For all details and more results the reader should consult \cite{DeSt,Desc}.
Still, there are many open questions in this approach, e.g. the
problem of $\pi^0-\eta$ mixing has so far not been solved.

\section{Reordering the chiral expansion II: Heavy kaons}
\def\theequation{\arabic{section}.\arabic{equation}}
\setcounter{equation}{0}
\label{sec:heavyK}

\begin{figure}[htb]
\centerline{
\epsfysize=6cm
\epsffile{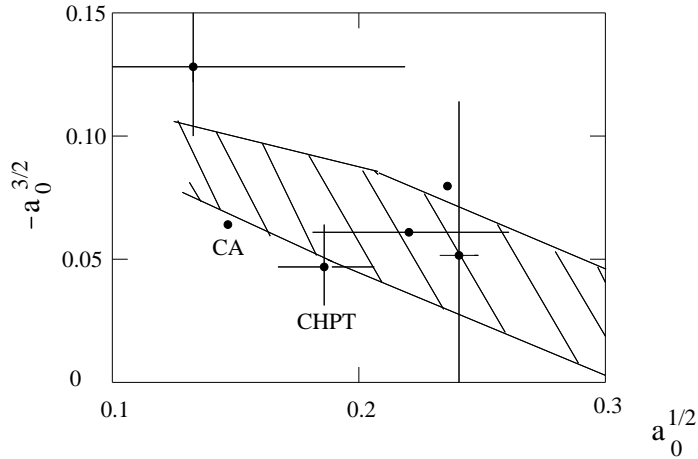}
}
\centerline{\parbox{10cm}
{\caption{Chiral predictions for the S-wave $\pi K$ scattering lengths
    in pion mass units. Lowest order: CA, one-loop: CHPT. The shaded
    area are the allowed values from the old Roy  equation study of
    \protect\cite{JoNi}. The data points can be traced back from 
    \protect\cite{BKMpik}.
         \label{fig:pik}}}}        
\end{figure}

Let me start this section by briefly reviewing some results concerning
low-energy pion-kaon scattering, which is the simplest Goldstone boson scattering
process involving strange quarks. The one-loop representation has been
given and analyzed in \cite{BKMpik}. The chiral predictions for the
scattering lengths in the physical isospin channels are displayed in
Fig.~\ref{fig:pik} in comparison to the at that time available data
and constraints from dispersion theory (for the normalization $F_K = F_\pi$).  
\noindent
As has been stressed recently \cite{KuMe2}, it is advantageous to
consider the isospin-even and -odd scattering lengths, $a_0^+ =
(a^{1/2}_0 + 2 a^{3/2}_0)/3$ and $a_0^- = (a^{1/2}_0 - a^{3/2}_0)/3$,
respectively. The chiral analysis of these two quantities displays a
remarkably different behavior: As was pointed out e.g.\ in~\cite{AnBu}, 
$a_0^-$ at order ${\cal O}(p^4)$ 
depends only on one single low--energy constant, $L_5$, which is in
turn determined by
the ratio $F_K/F_\pi$, such that $a_0^-$ can be predicted to a 
very good accuracy, $a_0^- = (0.0793 \pm 0.0006) M_\pi^{-1}$. On the
contrary, no less than seven low--energy
constants enter the isoscalar scattering length $a_0^+$, some of them
known to rather poor
accuracy, such that the chiral prediction for this quantity at 
one--loop order is plagued by a very large uncertainty, 
$a_0^+ = (0.025 \pm 0.017) M_\pi^{-1}$.
Furthermore, the tree--level result for $a_0^-$ receives only a 12\% correction 
from one--loop contributions and counterterms (\emph{if}, as done here, one normalizes
the tree level result to $1/F_\pi^2$), 
while $a_0^+$ vanishes at tree level, and the contributions at order 
${\cal O}(p^4)$ are rather large. In fact, if one expands the 
expressions for $a_0^{+}$ and $a_0^{-}$ in powers of $M_\pi/M_K$, 
$a_0^-$ receives contributions of odd powers of the pion mass only, while
$a_0^+$ scales with even powers of $M_\pi$, such that one has symbolically:
\beqa
a_0^- &=& \frac{M_KM_\pi}{8\pi F_\pi^2(M_K+M_\pi)} \left\{
1+ \frac{M_\pi^2}{\Lambda_\chi^2} \left( c_0^- + c_2^- \frac{M_\pi^2}{M_K^2} 
+ c_4^- \frac{M_\pi^4}{M_K^4} + \ldots \right)
+ {\cal O}(\Lambda_\chi^{-4}) \right\} ~,
\\
a_0^+ &=& \frac{M_KM_\pi}{8\pi F_\pi^2(M_K+M_\pi)} \left\{
\frac{M_\pi M_K}{\Lambda_\chi^2} \left( c_0^+ + c_2^+ \frac{M_\pi^2}{M_K^2} 
+ c_4^+ \frac{M_\pi^4}{M_K^4} + \ldots \right)
+ {\cal O}(\Lambda_\chi^{-4}) \right\} ~,
\eeqa
where $\Lambda_\chi = 4\pi F_\pi$ is the chiral symmetry breaking scale,
and the $c_i^\pm$ are quark mass independent constants.
This makes it obvious that the one--loop contributions to $a_0^-$ 
are suppressed by one power of $M_\pi/M_K$ with respect to those to $a_0^+$.
This behavior is completely analogous to what one finds for pion--nucleon scattering 
lengths (see e.g.~\cite{BKMa-}). Therefore, it might pay to reorder
the chiral expansion by treating the kaons as massive (matter)
particles, as will be discussed in the next paragraph. First, however,
let me mention that progress has been made in connecting the
dispersive and chiral representations of the $\pi K$ scattering
amplitude, see \cite{AnBu}. This paves the way for a new dispersive
study of pion-kaon scattering. As a first application, sum rules for
certain LECs where considered in \cite{ABM}, and a new value for the
OZI-violating coupling $L_4$ was reported,
\beq 
L_4 (M_\rho ) = (0.22 \pm 0.30) \cdot 10^{-3}~,
\eeq
which is in its central value very different from the standard
one (see e.g. \cite{daphne}) but consistent within error bars.
Another interesting work concerning S-wave $\pi K$ scattering including
also explicit resonance fields was reported in \cite{JOP2} (extending
and improving upon the work of \cite{BKMres}). Isospin violation 
of relevance for the measurement of $\pi K$ atoms at CERN \cite{DIRAC}
was considered in  \cite{KuMe1,KuMe2,NeTa,Ne}.

\medskip\noindent
So far, we have considered the kaons and the pions on equal footing, namely
as pseudo-Goldstone bosons of the spontaneously broken chiral symmetry of QCD,
with their finite masses related to the non-vanishing current quark masses.
However, the fact that the kaons (and also the eta) are much heavier than the
pions might raise the question whether a perturbative treatment in the
strange quark mass is justified? Also, we had just seen in the discussion of
the chiral expansion of the S-wave scattering lengths that the kaon behaves
much like the nucleon, which is a genuine matter field. In fact, one can take 
a very different view from the standard case 
and consider only the pions as light with the kaons behaving as heavy sources, 
much like a conventional matter field in baryon CHPT. This point of view was 
first considered in the Skyrme model \cite{CK} and has been reformulated in the context of
heavy kaon chiral perturbation theory (HKCHPT) in Ref.\cite{Roessl} (a closely related
work applying reparametrization invariance instead of the reduction of 
relativistic amplitudes was presented in \cite{Oul}.). Since the kaons appear
now as matter fields, the chiral Lagrangian for pion-kaon interactions decomposes
into a string of terms with a fixed number of kaon fields, 
that is into sector with $n$ ($n \geq 0$)  in-coming and $n$ out-going kaons,
\begin{equation}
{\cal L}_{\rm HKCHPT}={\cal L}_{ \pi}+ {\cal L}_{\pi KK}+ 
{\cal L}_{\pi KK K K}+\ldots~.
\end{equation}
where the first term is the conventional pion effective Lagrangian.
To be specific, consider only processes with at most one
kaon in the in/out states (like e.g. $\pi K$ scattering). 
The general form of the Lagrangian up to one--loop accuracy, i.e. order 
${\cal O}(p^4)$,  is 
\begin{equation}
{\cal L}_{HK \chi PT}={\cal L}_{ \pi}^{(2)}+{\cal L}_{ \pi}^{(4)}+ 
{\cal L}_{\pi KK}^{(1)}+ {\cal L}_{\pi KK}^{(2)} + 
{\cal L}_{\pi KK}^{(3)}+ {\cal L}_{\pi KK}^{(4)}~,
\label{hklag}
\end{equation}
where the terms ${\cal L}_{\pi KK}^{(n)}$ with $n \ge 2$ contain 
so-called heavy kaon LECs, denoted $A_i$, $B_i$ and $C_i$, respectively.
Obviously, the power counting has to be modified due
to the new large mass scale, $M_K$; and as it is the case for baryons, terms
with an odd number of derivatives are allowed. 
The pertinent power counting rules  can be easily derived, similar
to the case of heavy baryon CHPT \cite{JeMe,BKMrev}.  Consider the amplitude 
${\cal A}$ of an arbitrary graph consisting of $V_n^{\pi \pi}$ pionic vertices 
of order  $n\,$, $V_m^{\pi K}$ pion--kaon vertices of order $m\,$, $E^{\pi }$ external 
pion legs, $E^{K }$ external kaon lines, $I^{\pi}$ internal pion lines, 
$I^{K}$ internal kaon lines, and $L$ loops. The chiral dimension $\nu$ 
assigned to such a diagram is (that is, ${\cal A} \sim p^\nu$)
\begin{equation}
\nu=\sum_n V_n^{\pi \pi}(n-2)+\sum_m V_m^{\pi K}(m-1)+2L+1~,
\label{fpower}
\end{equation}
where we have used the topological identities $I^K=\sum_m V_m^{\pi K}-1$
and $I^{\pi}+I^K=L+\sum_n V_n^{\pi \pi}+\sum_m V_m^{\pi K}-1\,$.
To provide HKCHPT with predictive power we need the numerical values 
of the renormalized LECs characteristic of the heavy kaon theory.
In principle, these can be obtained from experimental data, in complete analogy to the
determination of  the $L_i\,$ in conventional SU(3) CHPT. In fact, one can translate 
knowledge of the $L_i$ into the heavy kaon theory and thus infer information about the 
HKCHPT parameters. As already stressed, the major difference in both approaches is the 
treatment of the strange quark 
mass.  While  in SU(3) CHPT $m_s$ serves as an expansion parameter of the chiral series, 
in the heavy  kaon approach $m_s$ does enter as part of the static kaon mass, yet, 
when involved in loops, the kaon is rather dealt with as a heavy quark, i.e. its effects are 
absorbed into the numerical  values of the constants present in any expansion. Having 
calculated an observable quantity in both schemes, a comparison of the two power series 
then yields an expansion of certain combinations 
of HKCHPT parameters in powers of $m_s\,$, where the SU(3) CHPT LECs are incorporated 
into the coefficients. One can thus compare these two series order by order 
and relate the various LECs in both schemes. This procedure is referred to as matching.
For example, the comparison of the quark mass expansion  of
$M_\pi$, $F_\pi$ and $M_K$ in both scheme gives
one matching condition for some second order HKCHPT LECs,
\beq
A_1 + \frac{1}{4}\,  M^2 A_2 =  \frac{1}{2} + {\cal O} (\bar{M}_K^2)~,
\eeq
with $\bar{M}_K^2 = B_0 m_s$ and $M^2$ the leading (quark mass independent)
term in the chiral expansion of the kaon mass in the heavy kaon scheme.
Another relation is found from the fourth order corrections,
\beqa
&& {F^2 \over 8} \left( C_5^r + 2C_6^r \right) + F^2 \left( C_{13}^r +
  C_{14}^r +  C_{15}^r  \right) + {3M^2 A_2 \over 2048 \pi^2} \nonumber\\
&& \qquad =
-{5 \over 9216\pi^2} - {1 \over 2304\pi^2}
\log \left( {4\bar{M}_K^2 \over 3\mu^2}\right)
+ \frac{1}{4}\left(4L_4^r + L_5^r -8L_6^r+2L_8^r \right)
+ {\cal O} (\bar{M}_K^2)~, 
\eeqa
with $\mu$ the scale of dimensional regularization and we have suppressed the
scale dependence of the renormalized LECs (note that the dimension two LECs
are finite). This approach is particularly  suited to analyze chiral SU(2) 
theorems within a three flavor formulation. Pion-kaon scattering in the
threshold region was considered in Ref.~\cite{Roessl} and all  LECs appearing
in the scattering amplitude were obtained by matching conditions. This allows
e.g. to calculate threshold parameters (which mostly agree with the ones
obtained in \cite{BKMpik}).\footnote{Note, however, that the $\pi K$ amplitude as
published in \cite{Roessl} is not free of errors, see \cite{FKM}.}
Most interestingly, a low-energy theorem for the quantity $A^-$, 
\beq
A^- = a_0^- - 6M_\pi M_K \, a_1^- + 30 M_\pi^2 M_K^2 \, a_2^- ~, 
\eeq
first considered in Ref.~\cite{Lang}, was given. To leading order, $A^-_{\rm CA} =0$,
and the higher order contact terms and loops generate a small one--loop correction,
\beq
A^-_{\rm 1-loop} = {M_\pi \over M_\pi+M_K}{M_K^3 \over 160\pi^3 F_\pi^4} +
 {M_\pi^3 \over M_\pi+M_K}\left( {3M_K \over 32\pi^3 F_\pi^4} + {3B_2^r M_K^3
\over 2\pi F_\pi^2} + {3M_K \bar{\mu}_\pi \over \pi F_\pi^4} \right)
= 0.022~{M_\pi}^{-1}~.
\eeq 
More accurate data on the threshold parameters
would be needed to quantitatively test this prediction.
Other aspects of heavy kaon CHPT are discussed in \cite{Oul} and in \cite{FKM}.
It would certainly also be of great interest to apply this scheme to
reactions involving eta mesons.

\section{Update on sigma terms}
\def\theequation{\arabic{section}.\arabic{equation}}
\setcounter{equation}{0}
\label{sec:sigma}

In QCD, the mass terms for the three light quarks
$u$,$d$, and $s$ can be measured in  the so--called sigma--terms.
These are matrix elements of the scalar quark currents $m_q \bar{q}q$  
in a given hadron $H$, $\langle H| m_q \bar{q}q | H \rangle$, with $H$
e.g.  pions, kaons or nucleons. Since no external scalar probes
are available, the determination of these matrix elements proceeds
by analyzing four--point functions, more precisely Goldstone boson--hadron 
scattering amplitudes in the unphysical region,
$\phi (q) + H(p) \to \phi (q') + H (p')$ (note that the hadron can
also be a Goldstone boson).  The determination of the sigma--terms starts from
the generic  low-energy theorem \cite{BPP} 
for the isoscalar scattering  amplitude $A(\nu,t)$\footnote{We use
the standard Mandelstam variables $s,t,u$ (subject to the constraint
$s+t+u = 2M_\phi^2 + 2M_H^2$) to describe the scattering process
and introduce the crossing variable $\nu = s-u$.}
\beq\label{LET}
F^2 A(t,\nu) = \Gamma (t) + q'^\mu q^\nu \, r_{\mu\nu}~,
\eeq
where $\Gamma (t)$ is the pertinent scalar form factor (of course, one
has to differentiate between non-strange ($\sim \bar u u + \bar d d$ and
strange ($\sim \bar s s$) form factors, see e.g. Eq.~(\ref{sff}))
\beq
 \Gamma (t) = \langle H(p') \, | \, {m_q} \bar q q \, | \, H (p) \rangle ~, ~ 
t = (p'-p)^2~,
\eeq
which at zero momentum transfer gives the desired sigma term,
\beq
\Gamma (0) = 2 M_\phi \, \sigma_{\phi H}~,
\eeq
for appropriately normalized hadron states. Furthermore, in Eq.(\ref{LET})
$r_{\mu\nu}$ is the so--called remainder, which is not determined by
chiral symmetry. However, it has the same analytical structure as the
scattering amplitude. To determine the sigma--term, one has to work in a
kinematic region where this remainder is small, otherwise a precise determination
is not possible. By definition, the point where the remainder takes its
smallest value is the so--called Cheng-Dashen (CD) point~\cite{cd}, which e.g. for pion
scattering off other hadrons is given by $t = 2M_\pi^2$ and $\nu = 0\,$,
which clearly lies outside the physical region for elastic scattering
but well inside the Lehmann ellipse.

\medskip\noindent
Consider first the case of pion-pion scattering, which only involves
light quarks. Here, the situation is under complete control, we just
quote the result for the LET, Eq.~(\ref{LET}), from the work of Ref.~\cite{GS}.
At tree level, the remainder is zero at the CD-point (but sizeable at threshold
$t=0\,$!) and to one-loop accuracy one has
\beq\label{spion}
\begin{tabular}{ccccc}
1.14 & = & 1.09 & + & 0.05\phantom{~,} \\
$F^2_\pi A_{\pi}^{\rm CD}$ & = & $\Gamma_{\pi} (2M_\pi^2)$  & 
+ & $\Delta_{\pi}$~,\\
\end{tabular}
\eeq
in pion mass units and $\Delta_\pi$  is the remainder at the CD-point. 
For the pion sigma term it amounts to a correction of about 3.5$\,$MeV
with the total pion sigma term at one loop being $\sigma_\pi \sim 69\,$MeV
\cite{GL1}. Of course, it is known that there are large higher order
corrections in this channel. Their influence on the one-loop result 
Eq.~(\ref{spion}) could be worked out using the existing two-loop or
dispersive representations of the scalar pion form factor and the $\pi\pi$
scattering amplitude. Astonishingly, for the case of pion-kaon scattering,
matters are not very different at one-loop order \cite{FKM} (scaled by
a factor of two for easier comparison with the pion case)
\beq\label{skaon}
\begin{tabular}{ccccc}
1.09 & = & 1.07 & + & 0.02\phantom{~,} \\
$F^2 A_{\pi K}^{\rm CD}$ & = & $\Gamma_{K} (2M_\pi^2)$  
& + & $\Delta_{\pi K}$~,\\
\end{tabular}
\eeq
again in pion mass units and for the case that one choses $F^2 = F_\pi^2$ for the
field normalization. Furthermore, $\Gamma_K$ is the non-strange scalar
form factor of the kaon, defined here via
\beq
 \Gamma_{K} (t) = \langle K^0(p') \, | \, \hat{m} (\bar u u + 
 \bar d d) \, | \, K^0 (p) \rangle ~.
\eeq
In three flavor CHPT, up to corrections of higher
order, one could also set $F^2 = F_\pi F_K$ or $F^2 = F_K^2$. In these
latter two cases, the LET is afflicted by larger corrections $\sim L_5 M_K^2
M_\pi^2$. However, employing the heavy kaon formalism discussed in 
section~\ref{sec:heavyK}, the choice $F^2 = F_\pi^2$ has to be made (because
in this case the kaon cannot decay) and thus the LET, Eq.~(\ref{skaon}), can be considered
a true two-flavor theorem within a three flavor calculation. For a more detailed
discussion of this and related issues, the reader is referred to \cite{FKM}.
The corresponding $\sigma$--term is $\sigma_{\pi K} \simeq 36\,$MeV,
very similar to the pion-nucleon $\sigma$--term $\sigma_0$ (as discussed below).

\begin{figure}[H]
\begin{center}
\epsfysize=4.in
\epsfig{file=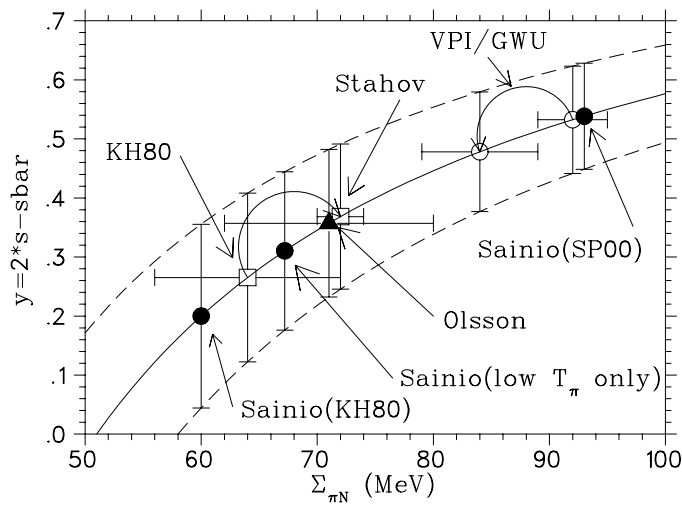, angle=90, width=3.5in}
\end{center}
\centerline{\parbox{11cm}
{\caption[gruel]{Values of the $\pi$N $\sigma$ term 
  (at the CD-point) from various recent analysis.
  The solid curve and its dashed counterparts indicate the
  relationship between $\sigma_{\pi N}$ and the $s\bar{s}$ content of
  the proton given in Ref.~\cite{BoMe} The points indicate the
  results of various analyses which lead to $\sigma_{\pi N}$.}
\label{fig:sigma}}}
\end{figure}

\medskip\noindent
The situation is more complicated for the pion-nucleon case, which is
certainly the most studied of all sigma terms. Chiral perturbation
theory can be used to calculate the remainder at the CD-point and to
relate the sigma term to the strangeness content of the proton. 
Consider first the remainder. While the scattering amplitude and the
scalar form factor both contain strong IR singularities, these cancel
completely at the CD-point up to and including terms of order $p^4$
\cite{BKMcd}, 
\beq
\Delta_{\pi N}^{\rm CD} = c M_\pi^4 + {\cal O}(M_\pi^5)~,
\eeq 
where $c$ is a quark mass independent
constant. Numerically, the remainder is small, $\Delta_{\pi N}^{\rm
CD} \sim 2\,$MeV. The relation between the sigma term, $\sigma_{\pi
N}(0)$, and the strangeness content, $y$, is \cite{Gasser}
\beq
\sigma_{\pi N} (0)  = {\sigma_0 \over 1-y}~, \quad y = {2 \langle
  p|\bar{s}s|p\rangle \over \langle  p|\bar{u}u + \bar{d}d|p\rangle}~.
\eeq
The constant $\sigma_0$ can be calculated in
baryon CHPT. Its latest evaluation at ${\cal O}(m_q^2)$ is $\sigma_0 =
(36\pm 7)\,$MeV \cite{BoMe}. 
To obtain the isospin-even D-amplitude of
pion-nucleon scattering with the pseudo-vector Born terms subtracted 
at the CD-point (i.e. the left-hand-side of Eq.~(\ref{LET})) from
$\pi N$ scattering data, one has to employ dispersion relations.
This field is under active reinvestigation by various groups since
new low-energy data from TRIUMF and PSI have become available and some
older data are now considered incorrect. Also, the commonly applied 
electromagnetic corrections have recently shown to be incomplete in
the low-energy region \cite{FeMe1}. The impact of that work on the
extraction of the hadronic amplitudes using dispersion relations
remains to be worked out. For other recent work on $\pi N$ scattering and
the sigma term in the framework of chiral perturbation theory, see
e.g. \cite{FeMe2,BuMe,BeLe1,BeLe2} (and references therein).
In Fig.~\ref{fig:sigma} most of the recent determinations of 
$\sigma_{\pi N}^{\rm CD} \equiv \Sigma_{\pi N}$ are shown in relation to the
strangeness content $y$ based on the work of \cite{BoMe} and
employing the scalar form factor from \cite{GLS}, $\sigma_{\pi N}
(2M_\pi^2) -\sigma_{\pi N} (0) \simeq 15\,$MeV.  The figure is taken
from \cite{MeSm}. The value of the GWU group has slightly changed,
their most recent number \cite{PASW} is $\Sigma_{\pi N} = (79\pm 7)\,$MeV, which
implies a huge strangeness content. Only when the new dispersion theoretical
analysis under development at Karlsruhe and Helsinki will have been finished,
more definite statements can be made. Note, however, that the most constrained 
and most internally consistent $\pi N$ $\sigma$ term analysis is the one of
\cite{GLS}, which does not lead to any absurdly large strangeness fraction.
One might also speculate that the scenario of a very suppressed three--flavor 
condensate might lead to a consistent picture of $\sigma_{\pi N}$ and the 
strangeness content of the proton.

\noindent

\section{Remarks and prejudices on light scalar mesons}
\def\theequation{\arabic{section}.\arabic{equation}}
\setcounter{equation}{0}
\label{sec:scalar}

\begin{figure}[htb]
\parbox{.48\textwidth}{\epsfig{file= 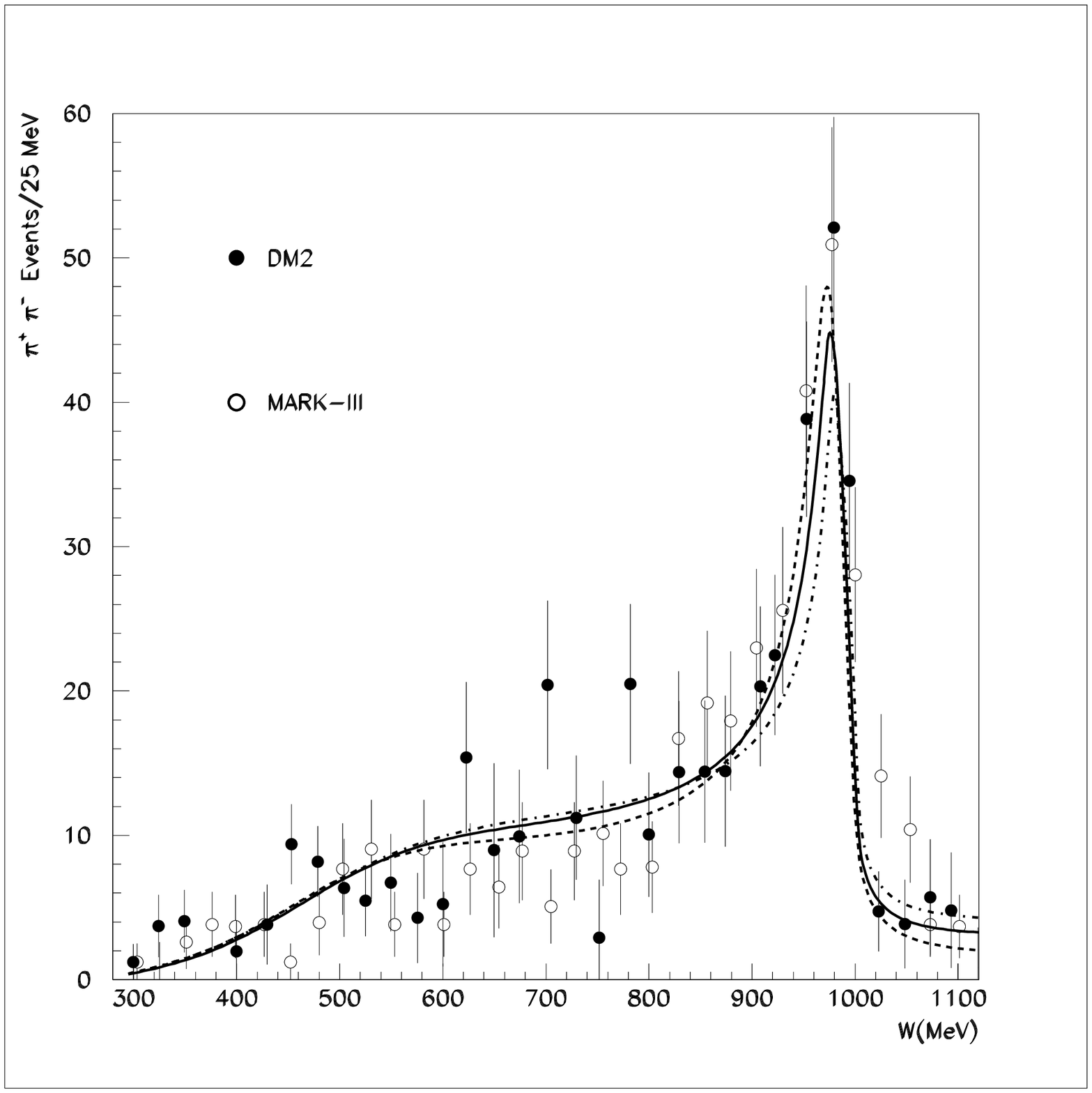,width=.46\textwidth,silent=,clip=}}
\hfill
\parbox{.48\textwidth}{\epsfig{file= 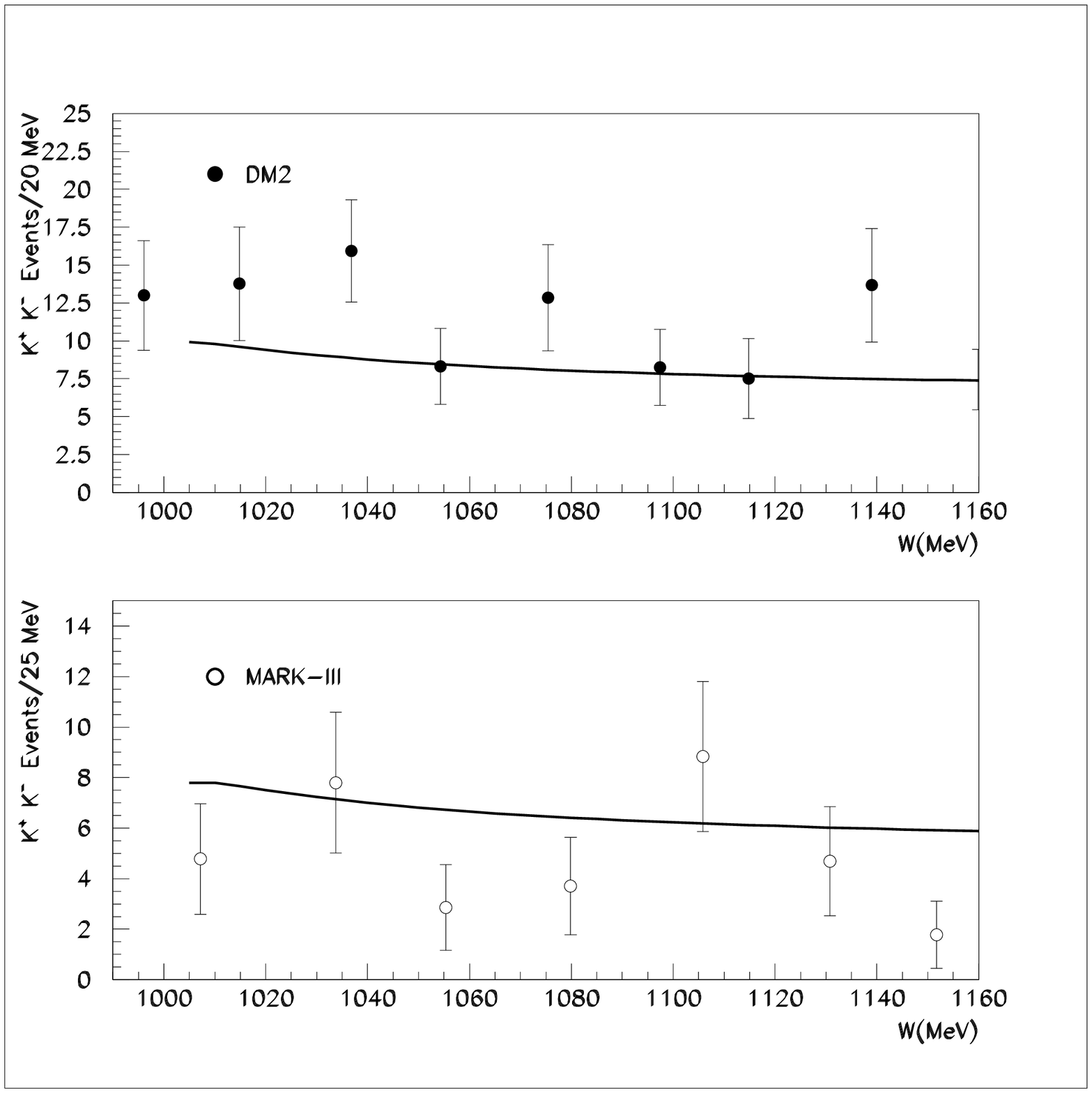,width=.46\textwidth,silent=,clip=}}
\smallskip

\caption{\label{fig:JPsi}
Left panel: $\pi^+\pi^-$ event distribution in the $J/\psi \to
\phi\pi^+\pi^-$ decay from  \protect \cite{MeOl}. The width of the bin is
25~MeV. The dashed lines indicate some theoretical uncertainty, 
see \protect \cite{MeOl}. Right panel: 
$K^+K^-$ event distribution in the $J/\Psi \rightarrow \phi K^+ K^-$ decay. 
The upper panel corresponds to the data from DM2 \cite{dm2}, 20 MeV
bins. The lower one corresponds to the data from MARK-III \cite{mk3}, 25 MeV bins.
We are grateful to David Bugg for useful discussion on these data.
}
\end{figure}

The nature of the low--lying scalar mesons has been and still is under
active debate.  This controversy originates from the observation that   
there are several different models to deal with the isospin $I=0 , \,1$ scalar   
sector, all of them reproducing the scattering data up to some extent, but with 
different conclusions with respect to the origin of the underlying
dynamics. Models for these scalar mesons include $\bar q \bar q qq$
states, mixing with glueballs, kaon-antikaon molecules and so on 
(for a recent review, see e.g. \cite{Penn}).\footnote{Note in particular
the closeness of the $f_0$ and $a_0$ to the $K\bar{K}$ threshold.}
To my opinion, these states are generated by the strong
final state interactions in the coupled $\pi\pi / K\bar{K}$ system,
for a pedagogical discussion see Ref.\cite{Ulfcnpp}. Of course, this
has to be backed by precise calculations. In particular, it is not
sufficient to describe the mass spectrum, but one also has to be able to 
properly account for strong and electromagnetic decays as well as 
production processes, in which these states dominate the final state
interactions. An attractive feature of such a dynamical generation of
the scalars (with vacuum quantum numbers) via loop effects
(rescattering) is the natural emergence of OZI violation in this channel, 
see e.g. \cite{Jan,IsGe}. With respect to this controversy, the
contribution of the work presented in \cite{OO1,OO2} is of importance.
In that approach, the most  
general structure of a partial wave amplitude when the unphysical cuts are neglected was  
established. In particular, in \cite{OO2}  explicit s-channel  
resonance exchanges were included together with the lowest order CHPT
contribution and the whole SU(3)  scalar sector with isospin
$I=0,1/2,1$ was studied.  It was observed that the  
lightest $0^{++}$ nonet is of dynamical origin, i.e. made up of meson--meson  
resonances, and is formed by the $\sigma(500)$,   
$\kappa$, $a_0(980)$ and a strong contribution to the physical $f_0(980)$. On the other  
hand, the pre--existing scalar nonet would be made up by an octet around 1.4 GeV and a  
singlet contributing to the physical $f_0(980)$ resonance. Here, pre--existing is to
be understood in the sense that the state under consideration {\em cannot} be explained
entirely in terms of rescattering, in the language of field theory an explicit matter
field representation is mandatory. The inclusion of a pre--existing contribution to the  
$f_0(980)$ was considered in order to be able to  
reproduce the data on the inelastic $\pi\pi\to K\bar{K}$ cross section
when   including also the $\eta\eta$ channel 
(note, however, that the errors on these data are under debate).
If this channel is not considered, one can   
reproduce the strong scattering data, including also the previous experiments on the  
inelastic $\pi\pi \rightarrow K\bar{K}$ cross section, without including such preexisting   
contribution. Finally, in Ref.\cite{OO2} it was also shown that the  
contributions in the physical region due to  the unphysical cut 
are very small.
As pointed out in \cite{AMP,MoPe}, to obtain a consistent picture of the scalar sector,  
one also has to study other reactions in which the $0^{++}$   
amplitudes have a possible large influence via final state  
interactions. In this way one can complement the deficient  
information coming from the direct strong S--wave  scattering data and distinguish between  
available models. In \cite{OO3}  the whole set of photon  
fusion reactions $\gamma \gamma \rightarrow \pi^0\pi^0$, $\pi^+ \pi^-$, $K^+ K^-$, $K^0   
\bar{K}^0$ and $\pi^0 \eta$ were reproduced in an unified way from threshold up to   
$\sqrt{s}\approx 1.4$. Furthermore, in
Ref.~\cite{MHOT} the reactions $\phi \to \gamma \pi^0 \pi^0$, $\gamma \pi^+ \pi^-$ 
and $\gamma   \pi^0 \eta$ were predicted. These predictions were nicely confirmed
by a recent experiment at Novosibirsk \cite{novo} (see also earlier
work in \cite{acha}).   
Finally, the already mentioned data on the OZI suppressed process $J/\Psi \to \phi \pi\pi
\,(\bar{K}K)$ could be described making use of the same coupled channel T-matrix in
\cite{MeOl}, see Fig.~\ref{fig:JPsi}.
\noindent
Within this  theoretical approach to the scalar sector,
one is thus able to discuss all these processes in an unified   
way. This is achieved without including new elements ad hoc for each  
reaction, because all these processes are related by the use of an  
effective theory description  that combines CHPT and unitarity constraints.  

\medskip\noindent
There is one other important topic which needs to be discussed in this
context. Irrespective of this admittedly controversial assignment of the
scalars, the pion scalar form factor {\em cannot} be represented simply in 
terms of a scalar meson with a certain mass and width.
Although the {\it description} of the low-lying scalar states
in terms of dynamically generated, rather than ``pre-existing,'' states may
be controversial and thus subject to ongoing discussion, let us stress
that the scalar form factor itself is not. The form factor which
emerges for the chiral unitary approach shown before, cf. Eq.~(\ref{sff}),
compares well  to that which emerges from the dispersion analysis of Ref.~\cite{DGL}.
\begin{figure}[tb]
\begin{center}
\epsfig{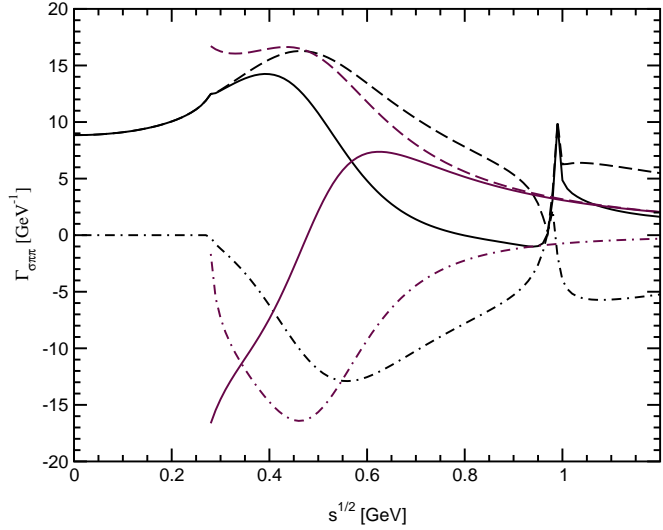}
\end{center}
\centerline{\parbox{10cm}{\caption{
The $\sigma\to\pi^+(p_+)\pi^-(p_-)$ form factor $\Gamma_{\sigma\pi\pi}$ as
a function of $\sqrt{s}$, with $s=(p_+ + p_-)^2$.
The real (solid line) and imaginary (dot-dashed line) parts of
$\Gamma_{\sigma\pi\pi}$, as well as its modulus (dashed line), are
shown. The curves which do not persist below physical threshold,
$\sqrt{s}=2M_{\pi} \sim 0.27$ GeV, correspond to the form factor
based on Eq.~(\ref{BWrun}), whereas the curves which extend to
$s=0$ correspond to the form factor from \cite{MeOl}.
\label{fig:sffcomp}
}}}
\end{figure}
\noindent
Consequently, when one has a source with vacuum quantum numbers coupled to a 
two-pion state, it can be very misleading to use a simple Breit-Wigner parameterization,
albeit with a running width. Specifically
\begin{equation}
\label{BWrun}
\Gamma_{\sigma  \pi\pi} (x) =
{g_{\sigma \pi^+  \pi^-} } \left(
\frac{1}{x-M_\sigma^2 + i\Gamma_\sigma (x) M_\sigma}
\right)\;,
\end{equation}
where the running width $\Gamma_\sigma(x)$ is defined as
\begin{equation}
\Gamma_\sigma (x) = \Gamma_\sigma \frac{M_\sigma}{\sqrt{x}}
\frac{\sqrt{x/4-M_\pi^2}}{\sqrt{M_\sigma^2/4 -M_\pi^2}} ~,
\end{equation}
in terms of the sigma meson mass $M_\sigma$ and its width $ \Gamma_\sigma$.
This is displayed graphically in Fig.~\ref{fig:sffcomp}, where the scalar
form factor is normalized such that in its peak it agrees with the Breit-Wigner
form with $M_\sigma=478\pm 24$ MeV and $\Gamma_\sigma=324\pm 41$ MeV as recently 
deduced from $D\to 3\pi$ decays \cite{e791ex},
\begin{equation}
\label{gammanorm}
\chi \, \left| \Gamma_1^n (M_\sigma^2) \right| =
\frac{g_{\sigma \pi^+ \pi^-}}{\Gamma_\sigma (M_\sigma^2) M_\sigma}~ \to
\chi=20.0\,\hbox{GeV}^{-1}~,
\end{equation}
using $g_{\sigma\pi\pi} = 2.52\,$GeV.
Note  that this significant difference casts doubt on the extraction
of the $\sigma$ meson properties from $D\to 3\pi$ decays, note
Ref.~\cite{e791ex}. This is quite in contrast to the pion vector form factor
and the $\rho$ meson, for which such a description works to 
good accuracy. It was shown in \cite{Sue} that the use of the proper
non-strange scalar pion form factor plays an important role in the description of
$B \to \rho \pi \to 3 \pi$  decay (extending and improving the work presented
in \cite{deA}). Moreover, the ``doubly'' OZI-violating form factor
$\langle 0 | \bar s s | \pi \pi\rangle$ is non-trivial as well, cf. Fig.~\ref{fig:OZI};
such a contribution is needed to fit the $\pi\pi$ and $K\bar K$ invariant mass
distributions in $J/\psi \to \phi \pi\pi (K\bar K)$ decay~\cite{MeOl}.
These observations give new insight on rescattering effects in
hadronic B decays, generating a new mechanism of factorization
breaking in $n\ge 3$ particle final states.

\section{``Strange'' baryons}
\def\theequation{\arabic{section}.\arabic{equation}}
\setcounter{equation}{0}
\label{sec:baryon}

As it is the case in the meson sector, the baryon spectrum also
reveals some ``strange'' states that might not
be genuine quark model states but dynamically generated
by strong final state interactions. The premier example is the
$\Lambda (1405)$. First speculations about its possible unconventional
nature date back to \cite{Dal}. Since then many (QCD-inspired) models
have been considered, but the first work of supplementing coupled
channel dynamics with chiral Lagrangians which allows to dynamically
generate the $\Lambda (1405)$ was reported in \cite{KSW},
see also \cite{OsRa}. A  non-perturbative resummation scheme is
mandatory  since in a perturbative theory
like CHPT, one can never generate a bound state or a resonance. There
exist many such approaches, but it is possible and mandatory to link 
such a scheme tightly
to the chiral QCD dynamics. Such an improved approach was
developed for pion--nucleon \cite{MeOl1} and later applied to $\bar K$N
scattering~\cite{MeOl2}. To be specific, let us consider $\pi$N
scattering. The starting point is the T--matrix for any partial wave,
which can be represented in closed form if one neglects for the moment
the crossed channel (left-hand) cuts (for  more explicit details, see
\cite{MeOl1})
\begin{equation}
T = \left[ \tilde{T} (W) + g(s) \right]^{-1}~,
\end{equation}
with $W = \sqrt{s}$ the cm energy (note that the analytical structure is much
simpler when using $W$ instead of $s$).
$\tilde{T}$ collects all local terms and poles (which can be
most easily  interpreted in the large $N_c$ world) and $g(s)$ is the
meson-baryon loop function (the fundamental bubble) that is resummed by e.g.
dispersion relations in a way to exactly recover the right-hand
(unitarity) cut contributions. The function $g(s)$ needs
regularization, this can be best done in terms of a subtracted
dispersion relation and using dimensional regularization (for details,
see \cite{MeOl1}). It is important to ensure that in the low-energy
region, the so constructed T--matrix agrees with the one of CHPT (matching).
In addition, one has to recover the contributions from the left-hand
cut. This can be achieved by a hierarchy of matching conditions, e.g.
for the $\pi$N system one has
\begin{eqnarray}
{\cal O}(p) &:& \tilde{T}_1 (W) = T_1^{\chi} (W) ~, \nonumber \\ 
{\cal O}(p^2) &:& \tilde{T}_1 (W) + \tilde{T}_2 (W) =  T_1^{\chi} (W)
+ T_2^{\chi} (W) ~, \nonumber \\
{\cal O}(p^3) &:& \tilde{T}_1 (W) + \tilde{T}_2 (W)+ \tilde{T}_3 (W) 
=  T_1^{\chi} (W)+ T_2^{\chi} (W) + T_3^{\chi} (W) +  \tilde{T}_1
(W) \, g(s) \,  \tilde{T}_1 (W)~,
\end{eqnarray}
and so on. Here, $T_n^{\chi}$ is the T--matrix calculated within
CHPT to ${\cal O}(p^n)$. Of course, one has to avoid double counting as
soon as one includes pion loops, this is achieved by the last term in
the third equation (loops only start at third order in this case).
In addition, one can also include resonance fields by saturating the
local contact terms in the effective Lagrangian through explicit meson and
baryon resonances (for details, see \cite{MeOl1}). In particular,
in this framework one can cleanly separate genuine quark resonances
from dynamically generated resonance--like states. The former require
the inclusion of an explicit field in the underlying Lagrangian, whereas
in the latter case the fit will arrange itself so that the couplings to 
such an explicit field will vanish.
This method was
applied to $\pi$N scattering below the inelastic thresholds in \cite{MeOl1} by
matching to the third order heavy baryon CHPT results and including
the $\Delta (1232)$, $N^\star (1440)$, $\rho (770)$ and a scalar
resonance. Instead of the CHPT low--energy constants (LECs), one now fits
resonance parameters, of course, to a given order one can only determine
as many (combinations) thereof as there are LECs.
\begin{figure}[ht]
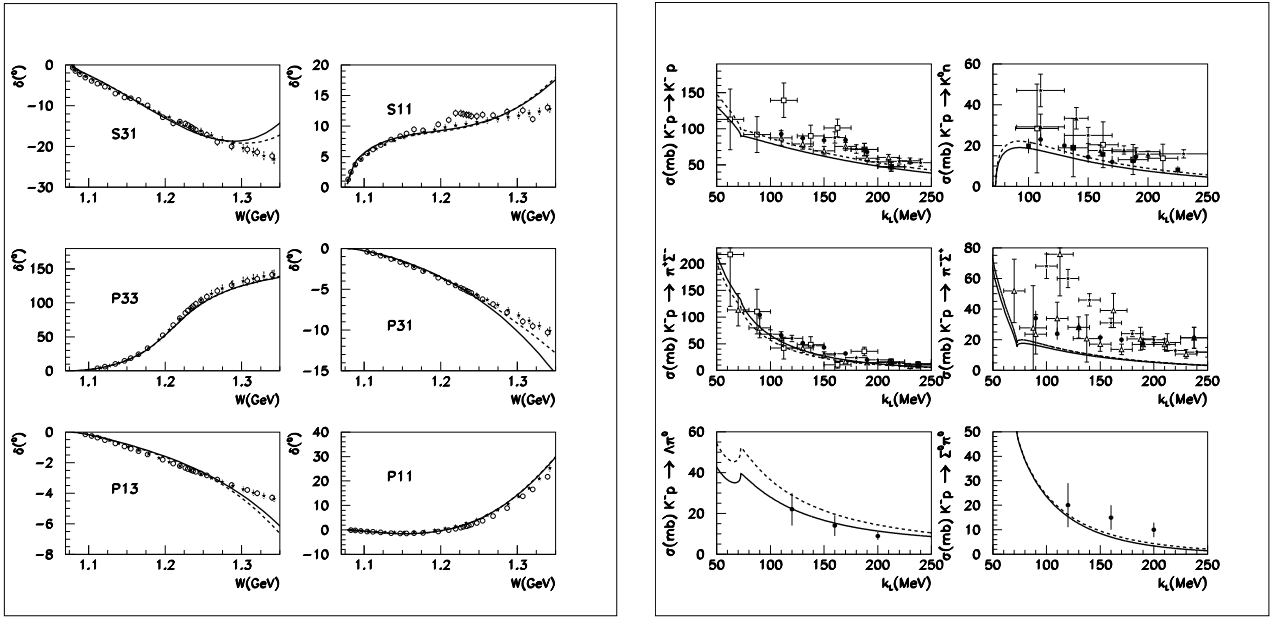

\parbox{.49\textwidth}{\epsfig{file= paw2.epsi,width=.48\textwidth,silent=,clip=}}
\hfill
\parbox{.49\textwidth}{\epsfig{file= scatf.epsi,width=.48\textwidth,silent=,clip=}}
\caption{\label{fig:fits}
Left panel: Fit to the low (S,P) $\pi$N partial waves. The solid (dashed)
lines refer to (un)constrained fits as explained in \cite{MeOl1}. Right
panel: Fit to various cross sections coupling to the $\bar K$N
channel. Solid lines: best fit, dashed lines: natural values for the
parameters, see \cite{MeOl2}.
}
\end{figure}
A typical fit to the low partial waves is shown in the
left panel of Fig.~\ref{fig:fits}. The threshold parameters are found
to be in good agreement with values obtained from phase shift analyses 
and the $\Delta$ is found in the complex--W
plane at (1210-i53)~MeV, in good agreement with earlier findings
\cite{delta}. It is also important to point out that the
scalar exchange can be well represented by contact terms, i.e. no need
for a light sigma meson arises.
These considerations were extended to
S--wave, strangeness $S = -1$ $\bar K$N scattering in \cite{MeOl2}. 
In this case, one has to consider the 
coupling to the whole set of SU(3) coupled channels, these are
$\bar K$N, $\Lambda \pi$, $\Sigma \pi$, $\Sigma \eta$ and $\Xi K$
(for earlier related work, see e.g. \cite{KSW,OsRa}). The lowest 
order (dimension one) effective Lagrangian was used, it depends on three
parameters, which are the average baryon octet mass, the pion decay
constant in the chiral limit and the subtraction constant appearing in
the dispersion relation for $g(s)$.
Their values can be estimated from simple considerations leading to the
so--called ``natural values''. 
One finds a good description of the
scattering data and the threshold ratios, see the dashed lines in the
right panel of Fig.~\ref{fig:fits}. Leaving these parameters free, one
obtains the best fit (solid lines). It is worth to stress that the
values of the parameters for the best fit differ at most by 15\% from their natural
values. We have also investigated the pole structure of the S--wave
$\bar K$N system in the unphysical Riemann sheets. In addition to the
$I = 0$ pole close to the $\bar K$N threshold that can be identified
with the $\Lambda (1405)$ resonance, one finds another pole with $I =
0$ close to the $\Sigma \pi$ threshold and another one with $I = 1$
close to the $\bar K$N channel opening (which is threefold degenerate
in the isospin limit). Thus one can speculate about a nonet of $ J =
1/2$ meson--baryon resonances with strangeness $S = -1$. Still, one has
to investigate the $I = 1/2$ channel with $S =0, -2$ in this energy
interval to strengthen this conjecture. In a similar approach, the
nature of the $S_{11} (1535)$ was investigated in \cite{KSW2}. Higher
resonances with $S=-1$ were considered in \cite{ORB}. To
differentiate the bound-state scenario from e.g. a three quark
description, it is mandatory to consider also decays and
electromagnetic properties, see e.g. \cite{NOTR} and \cite{FJSWS} for 
photo/electroproduction of the $\Lambda (1405)$ and the  $S_{11}
(1535)$, respectively.

\section{Strange goings-on in proton--proton collisions}
\def\theequation{\arabic{section}.\arabic{equation}}
\setcounter{equation}{0}
\label{sec:pp}

With the advent of Cooler Synchrotrons at IUCF, J\"ulich and Uppsala,
a huge amount of very precise data on meson production in
proton--proton collisions has become available and has led to many
theoretical investigations. For some general ideas, I refer to
\cite{KMS} and to various reviews, which have appeared over the last
few years \cite{ppreviews} (see also \cite{Colin}).
It is not my intention to discuss the role
of strangeness production in $pp$ collisions in general
(as revealed e.g. in the reactions $pp \to pKY$, with $Y$ a hyperon,
or $pp\to ppM$, with $M = \pi^0, \eta, \eta ', \phi, \ldots\,$)
but rather to concentrate on two specific reactions which allow to 
test the ideas behind the strong final state interactions in the meson
and the baryon sector discussed in the preceding two sections.

\begin{figure}[htb]
\parbox{.59\textwidth}{\epsfig{file= 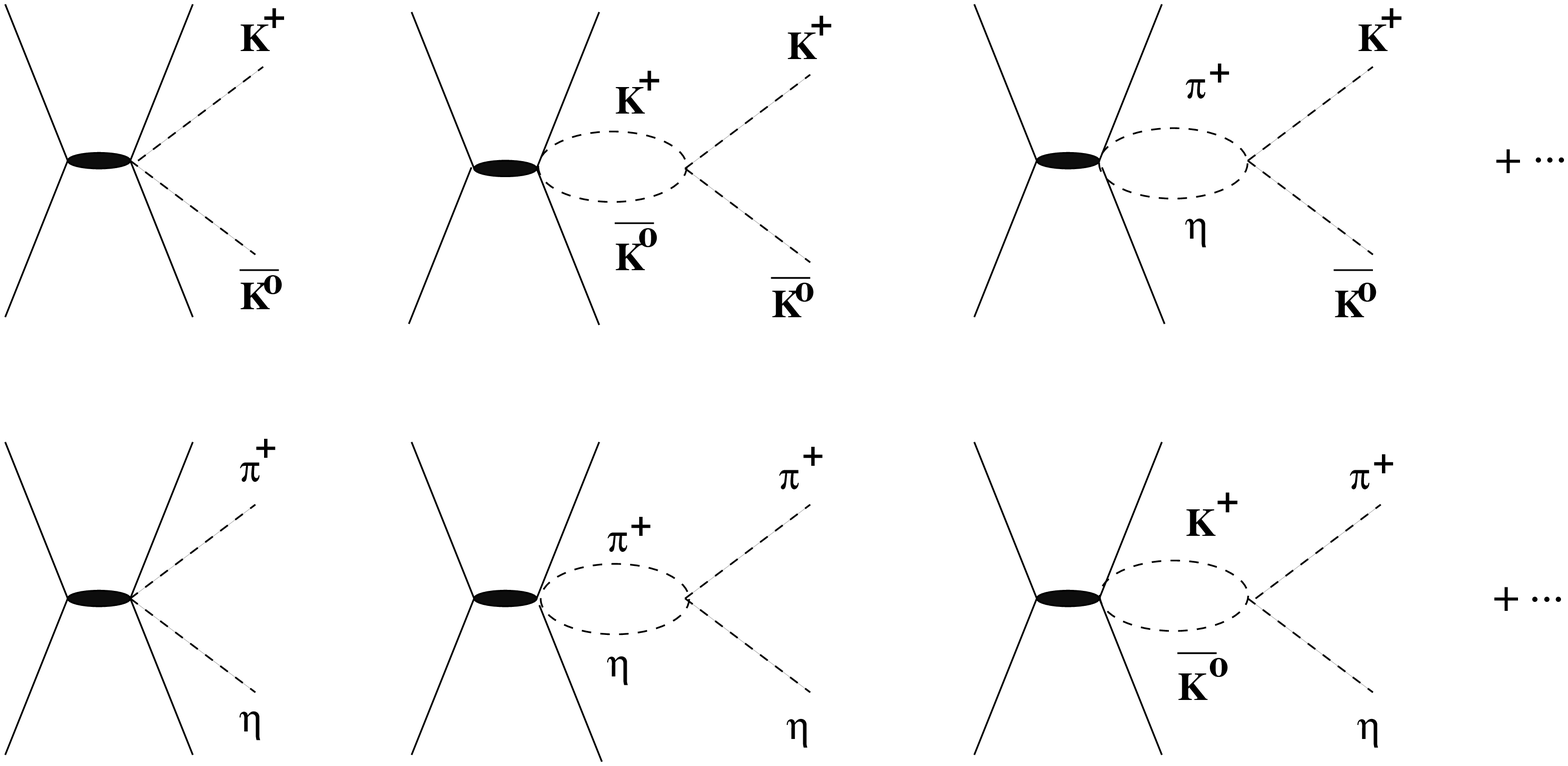,width=.57\textwidth,silent=,clip=}}
\hfill
\parbox{.39\textwidth}{\epsfig{file= 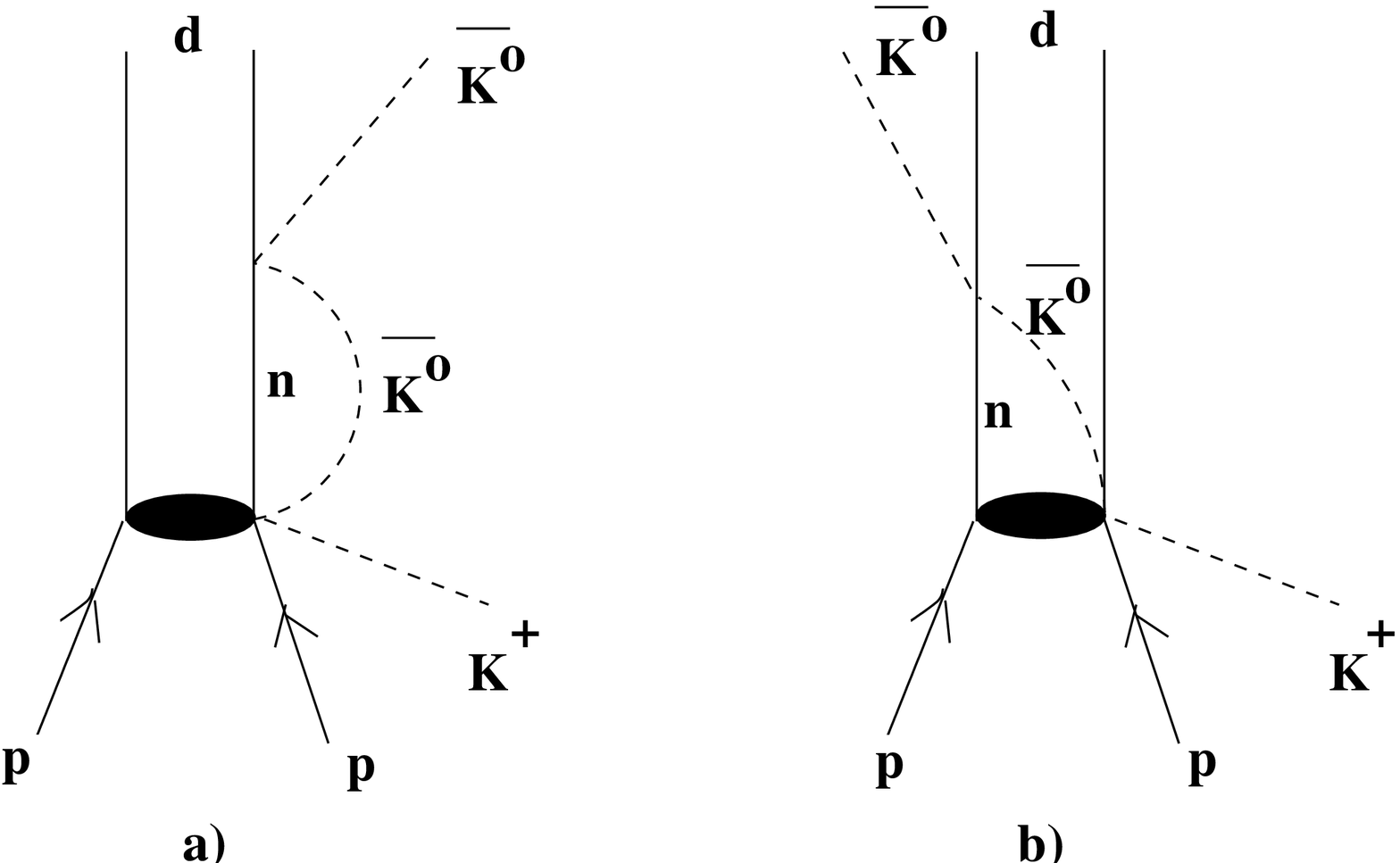,width=.37\textwidth,silent=,clip=}}
\smallskip

\caption{\label{fig:FSI}
Final state interactions in the coupled channel processes $pp \to d K^+
\bar{K}^0$  and $pp \to d \pi^+ \eta$.
Left panel: Meson--meson FSI that generate the $a_0 (980)$.
Right panel: Meson--baryon FSI sensitive to the $\Lambda (1405)$.
}
\end{figure}

\begin{figure}[tp]
\centerline{\epsfig{file=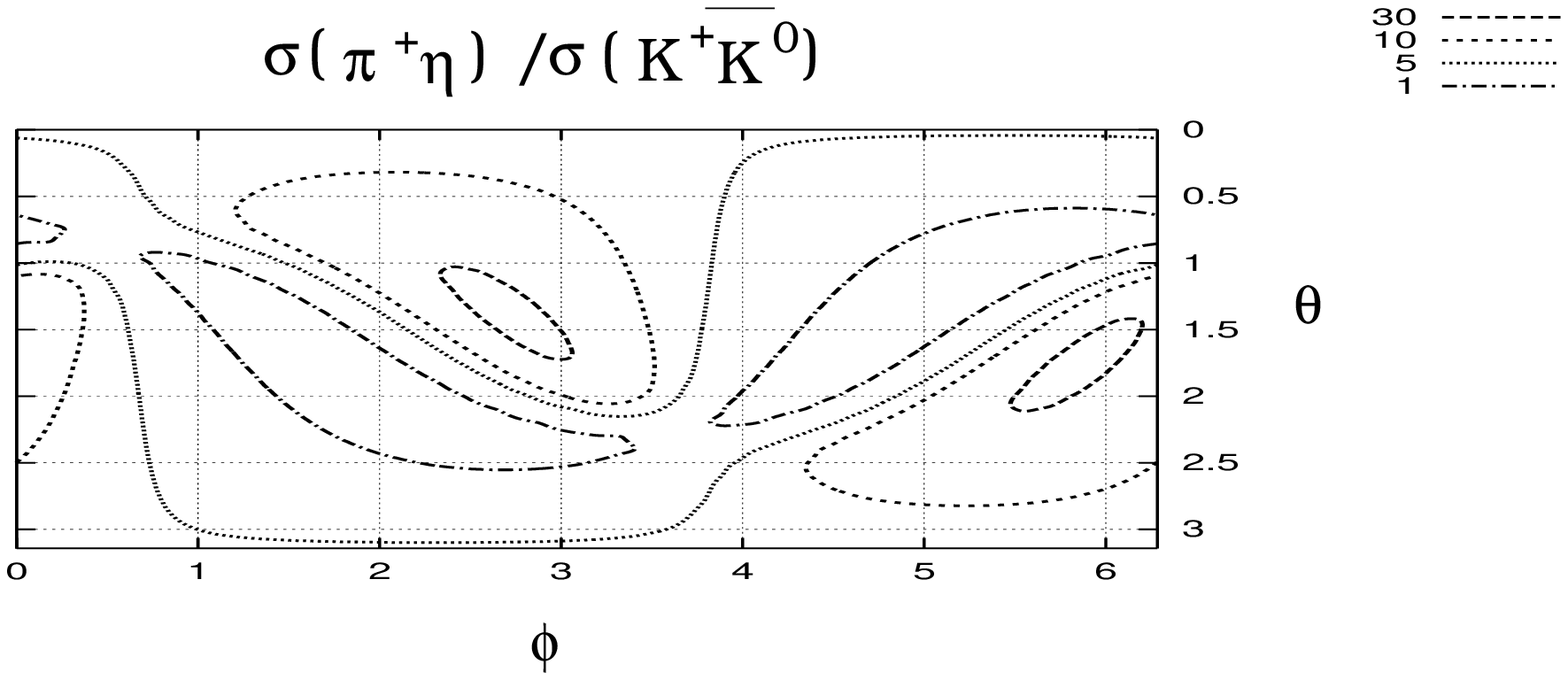,width=6.in}}
\vspace{-0.3cm}
\centerline{\parbox{12cm}
{\caption[pilf]{\protect \small $\sigma(\pi^+\eta)/\sigma(K^+\bar{K}^0)$ as function 
of $\theta$ and $\phi$ (in radians).
\label{fig:ccpe}}}}
\end{figure}

\medskip \noindent
Here I will concentrate on the (coupled channel) reactions 
$pp \to d K^+ \bar{K^0}$ and  $pp \to d \pi^+ \eta$ that are presently the 
subject of experimental study by the ANKE collaboration at 
COSY at J\"ulich with the aim (among others) of learning about the nature 
and properties of the $a_0(980)$ resonance \cite{anke}.
In the process $pp \to d K^+ \bar{K^0}$, the $K^+\bar{K}^0$ system is in an $I=1$ state 
which, given the proximity of the $a_0(980)$ resonance, would have its rate of 
production and invariant mass distributions very much influenced by the tail of 
that resonance. The reaction  $pp \to d \pi^+ \eta$
could actually display the  shape of the $a_0(980)$ resonance through 
the mass distribution of the $\pi^+\eta$ system.
The P-wave nature of these reactions \cite{Cass} is another peculiar 
feature that makes it different to other ones producing the 
$a_0(980)$. Indeed, due to total angular momentum and parity conservation 
as well as to the antisymmetry of the initial state, the two mesons cannot be 
simultaneously in intrinsic S-wave and in S-wave relative to the deuteron.
For the ANKE kinematics, one notes  that the kaon-antikaon pair is only 45~MeV
above threshold, thus the meson pair production is dominated by the operators
with $\ell =1, L=0$ and $\ell =0, L=1$, with $\ell$
the relative orbital angular momentum of the deuteron and the 
CM frame of the two pseudoscalars $PQ$ and  $L$ the orbital angular momentum 
of the latter in their own CM frame. Taking guidance from chiral symmetry 
\cite{BKMpieta}, we pointed out in \cite{OOM} that the structure $\ell =0, L=1$ 
is not realized for the $\pi^+ \eta$ final state and thus the primary production 
process can be parameterized in terms of three real amplitudes,
\beq\label{ppprod}
f_{\bar{K}K}^S = \cos \theta~, \quad f_{\pi\eta}^S = \sin \theta \cos \phi~, \quad 
f_{\bar{K}K}^P = \sin \theta \sin \phi~, 
\eeq
with $f_{\bar{K}K}^S$ the amplitude for producing the $\bar KK$ pair in an S-wave
and so on. Admittedly, this approach for the production mechanism is very simple,
but since it only contains two parameters (and one normalization), it has predictive
power. In \cite{OOM}, the general structure of the amplitudes for $pp \to d PQ$ was
also given which allows to apply the formalism of that paper to more complicated
models for the primary production. Quite in contrast to this, the important final
state interactions (FSI) are under complete control, employing the chiral unitary
approach already used in the previous sections. As shown in Fig.~\ref{fig:FSI},
these reactions have the unique feature of undergoing strong meson--meson as well
as meson--baryon FSI. 
\noindent The bubble sum in the coupled $\pi^+ \eta / \bar{K}^0 K^+$ system
generates the $a_0 (980)$, as described in section~\ref{sec:scalar}. On the other
hand, the strong $\bar K N$ interaction generates the $\Lambda (1405)$, which
leaves it trace in the large $\bar Kd$ scattering length, $a_{\bar Kd} = (-1.6 +
i 1.9)$~fm. The process at hand is indeed sensitive to the low--energy $\bar K N$ 
amplitude, see the right panel of  Fig.~\ref{fig:FSI}. Most importantly, these
two types of FSI can lead to very strong interference effects, which reveal
itself in the invariant mass distributions. Even more dramatic is the 
prediction for ratio
of the total $\pi^+ \eta$ to the $K^+\bar{K}^0$ production as a function
of the $\theta$ and $\phi$ parameters, see Eq.(\ref{ppprod}).
The  $\pi^+\eta$ production appears mostly
around the $a_0(980)$ resonance region. In Fig.\ref{fig:ccpe} we show the ratio 
between the integrated  $\pi^+\eta$ production cross section between 
$M_I=950$ MeV (the invariant mass of the meson pair) and the end of its 
phase space and the 
$K^+\bar{K}^0$ production cross section in all its available phase space. 
One sees that for most of the values of $\theta$ and $\phi$ 
the $\pi^+\eta$ production rate is substantially 
larger than that of $K^+\bar{K}^0$. It is interesting to point out that even in the 
case when there is no primary $\pi^+\eta$ production so that $f^S_{\pi\eta}=0$ 
($\theta=0$ and any $\phi$ or $\phi=\pi/2$, $3\pi/2$ and any $\theta$), the final state 
interactions starting from primary $K^+\bar{K}^0$ production can lead to a $\pi^+\eta$ 
cross section an order of magnitude bigger than that of the $K^+\bar{K}^0$. One also 
finds interesting interference effects for some values of $\theta$ and $\phi$ that can 
reinforce the $\pi^+\eta$ production as compared to the $K^+\bar{K}^0$ as well as 
other situations when the $K^+\bar{K}^0$ is produced more copiously than the 
$\pi^+\eta$ channel. The P--wave nature of this reaction can also be seen in 
the invariant mass distribution of the $\pi^+ \eta$ system; only
after division by the squared three--momentum of the deuteron, the resonance
shape of  the $a_0$ becomes visible. For a more detailed discussion and other
interesting predictions, see \cite{OOM}. Of course much more
theoretical effort is needed to achieve a truly systematic and
controlled description of these and similar hadronic meson production processes. 

\section{Summary}
\def\theequation{\arabic{section}.\arabic{equation}}
\setcounter{equation}{0}
\label{sec:summ}
Let me briefly summarize the salient topics discussed here:
\begin{itemize}
\item[$\bullet$]There are two distinct order parameters characterizing
the spontaneous chiral symmetry breaking in QCD, the pion decay constant in
the chiral limit, $F$, and the scalar quark condensate, $\Sigma$.
The condensate is more sensitive to the IR end of the spectrum of the Euclidean
Dirac operator. One expects these order parameters to decrease with increasing
number of massless (light) quarks.
\item[$\bullet$]Standard chiral perturbation theory which treats the
strange quark as light and is characterized by a large condensate
$\Sigma (3) \lesssim \Sigma(2)$ works in many, many instances. OZI
violation does not naturally emerge in this scheme, although  the OZI
suppressed low-energy constants could be large.
\item[$\bullet$]There are some indications that the three--flavor
condensate could be sizeably smaller than its two-flavor counterpart,
$\Sigma (3) \ll \Sigma(2)$ due to large quantum fluctuations related 
to light quark loops. In such a scenario, OZI violation would be natural and
be related to the sector of the light scalar mesons with vacuum quantum
numbers. However, there are still many open questions in this scenario.
\item[$\bullet$]Strange quarks can be treated as heavy matter fields, 
leading to a reordering of the chiral expansion. Matching conditions
relate this heavy kaon scheme to the standard formulation. The heavy
kaon approach can e.g. be used to study two--flavor low--energy theorems
within SU(3).
\item[$\bullet$]Sigma terms are a direct measure of the QCD quark mass term
within hadron states. The sigma terms related to pion--pion and pion--kaon
scattering are under control, whereas for the pion--nucleon case the situation
is less clear (mostly related to the analysis of $\pi$N scattering data).
\item[$\bullet$]The nature of the low--lying scalar mesons is under debate,
although a consistent picture emerges when these are considered as meson--meson
resonances due to strong final state interactions.  To the contrary,  the
scalar form factors of the pion and the kaon can be determined to good precision
and these can be further applied and tested in many different processes.
\item[$\bullet$]The bound state scenario for some  baryons like the
$\Lambda (1405)$  can be tested in hadronic and electromagnetic  
production processes or in transition form factors. 
Theoretical progress has been made in connecting the coupled
channel calculations to  chiral perturbation theory.
\item[$\bullet$]Goldstone boson pair production in proton--proton collisions
allows for further testing the final state interaction dynamics in the meson--meson
and meson--baryon systems. Clearly, the theory of these processes is only in its
infancy. 
\end{itemize}

\bigskip  

\section*{Acknowledgements}
I am  grateful to G\"oran F\"aldt for giving me the opportunity
to write up this contribution. All my collaborators, friends
and colleagues are warmly thanked for sharing their insight 
into the topics discussed here.

\end{document}